\documentclass[reprint, amsmath,amssymb, aps]{revtex4-2}
\usepackage{graphicx}
\usepackage{epstopdf}
\usepackage{dcolumn}
\usepackage{bm}
\usepackage{hyperref}
\usepackage{cleveref}
\usepackage{slashed}
\usepackage{amsmath}
\usepackage{amssymb}
\usepackage{xcolor}
\usepackage{footnote}

\begin{document}


\title{Properties of QCD axion in two-flavor color superconductive matter with massive quarks}

\author{Zhao Zhang}
\email{zhaozhang@pku.org.cn}
\author{Wenhao Zhao}

\affiliation{School of Mathematics and Physics, North China Electric Power University, Beijing 102206, China}

\begin{abstract}

We investigate the properties of QCD axion at low temperature and moderate density in the Nambu-Jona-Lasinio 
model with instanton induced interactions by simultaneously considering the scalar and pseudo-scalar condensates 
in both quark-antiquark and diquark channels. We derive the analytical dispersion relations of quarks with four-type 
condensates at nonzero theta angle $\theta=a/f_a$. The axion mass, quartic self-coupling, and the axion potential
are calculated in both the chiral symmetry breaking and two-flavor color superconducting phases. Using the commonly 
adopted model parameters, we find that due to the emergence of color superconductivity, the chiral phase transition 
not only does not lead to a significant decrease in axion mass and self-coupling, but rather results in an obvious 
enhancement of them. As a $\theta$ function, the axion potential exhibits an appropriate period of $\pi$, which is 
quite different from the case without considering the color superconductivity. The surface tension of axion domain 
wall is also calculated in the presence of color superconductivity.              

\end{abstract}


                              
\maketitle

\section{\label{sec:intro} INTRODUCTION}

The complicated nature of the QCD vacuum reveals that the effective Lagrangian should include an extra term
\begin{equation}
{\cal L}_{\theta_0}=\theta_0 \frac{g^2}{32\pi^2} G^a_{\mu\nu}\cdot\tilde{G}^{a\mu\nu}=\theta_0 Q, \label{eq:CPaa}
\end{equation}
where  $G^a_{\mu\nu}$ and $\tilde{G}^{a\mu\nu}$ denote the gluon field strength tensor and its dual respectively. 
$Q$ is the topological charge density and $\theta_0$ is a real parameter. The term \eqref{eq:CPaa} breaks both parity 
and time reversal symmetries but respects charge conjugation symmetry. Furthermore, taking into account the quark 
mass matrix $M_q$, the complete coefficient of $Q$ takes the form 
\begin{equation}
 {\theta}=\theta_0-arg(\text{det}M_q) \label{eq:CPaaa}.
\end{equation}
The phase angle $\theta$ is directly related to the neutron electric dipole moment (NEDM) $d_n$ via the 
predicted relation  
\begin{equation}
|d_n|\sim{10^{-16}}{\theta} e\cdot cm \label{eq:nEDM1}
\end{equation}
in the standard model \cite{Crewther:1979pi}. The experiment limit on the NEDM \cite{Baker:2006ts, Abel:2020pzs} 
suggests ${\theta} < {2}\times{10}^{-10}$. The extremely small ${\theta}$ indicates the CP (charge conjugation and parity) 
symmetry is conserved in strong interaction. This creates the so called fine-tuning or strong CP 
problem: the two sources of ${\theta}$ cancel with such precision is unnatural since they have distinct 
origins (one is related to the QCD vacuum and the other the Higgs mechanism). 

The Peccei-Quinn (PQ) mechanism is the most compelling solution to the strong CP problem which involves 
an extra $U(1)$ chiral symmetry beyond the standard model \cite{Peccei:1977hh,Peccei:1977ur}. The QCD axion 
is the pseudo-Nambu–Goldstone boson arising from the spontaneously breaking of $U(1)_{PQ}$ symmetry 
\cite{Weinberg:1977ma,Wilczek:1977pj} which may occur at the energy scale indicated by the axion 
decay constant $f_a$. The axion mass, self coupling, and couplings to other particles are all inversely 
proportional to $f_a$. Experimental constraints suggest that $f_a$ must be much higher than the electroweak 
breaking scale \cite{Kim:2008hd}. This implies QCD axion is a very weakly interacting particle with a 
small mass and thus is invisible \cite{Kim:1979if,Shifman:1979if,Dine:1981rt,Zhitnitsky:1980}. These properties 
make the axion one of the leading candidates for dark matter \cite{Sikivie:1982qv,Preskill:1982cy,Abbott:1982af,Dine:1982ah}. 

Besides QCD axion, the axion-like particles (ALP) are also proposed as the promising dark matter 
candidates\cite{Visinelli:2009zm,Duffy:2009ig}. As extremely light bosons, QCD axions and ALPs may form stars 
as well as the Bose-Einstein condensates \cite{Colpi:1986ye,Tkachev:1991ka,Kolb:1993zz,Chavanis:2011zi,Guzman:2006yc,Barranco:2010ib,Braaten:2015eeu,Davidson:2016uok,
Eby:2016cnq,Helfer2017black,Levkov:2016rkk,Eby:2017xrr,Visinelli:2017ooc,Chavanis:2016dab,Cotner:2016aaq,Bai:2016wpg,Sikivie:2009qn,Chavanis:2017loo}. 
The axions might be produced copiously in the interiors of stellar objects via the Primakoff process, 
the Compton-like process, axion bremsstrahlung, etc\cite{Caputo:2024oqc}. As a light and feebly interacting 
particle, the axion may influence the energy budget of stars drastically and affect the stellar evolution: 
axions may transport the energy of stars to outer space and shorten the star lifetime \cite{Sedrakian:2015krq,Sedrakian:2018kdm,Buschmann:2021juv,Leinson:2014ioa,Balkin:2022qer,Lopes:2022efy,Lucente:2020whw,Fischer:2021jfm}.
Moreover, a recent work has shown that the axion cloud may form around the neutron stars \cite{Noordhuis:2023wid}. 
Since we only study QCD axion in this paper, we refer to QCD axion as axion for convenience in the following. 

To understand the axion roles in cosmology and astrophysics, we must know the axion properties, e.g., 
axion mass, self-coupling, and couplings to normal matter in the hot and dense medium. 
The dependences of axion properties on the temperature and fermion chemical potentials determine 
how the axion affects the formation of large-scale structure of the universe, the cosmological evolution, 
and the properties and evolutions of stellar objects. In this work, we mainly concentrate 
on how the axion potential, axion mass and self-coupling change when QCD phase transitions 
happen at finite temperature and baryon number density. 

Most of properties of axion are determined by the non-perturbative QCD dynamics. In the literatures, 
the axion mass and quartic self-coupling in cool and hot medium had been studied using lattice QCD (LQCD)
\cite{Berkowitz:2015aua,Bonati:2015vqz,Borsanyi:2016ksw,Aoki:2017imx,Petreczky:2016vrs} and chiral perturbation theory($\chi\text{PT}$)
\cite{GrillidiCortona:2015jxo}. However, applying LQCD in dense medium has the limitation because of the sign 
problem and $\chi\text{PT}$ fails to describe QCD phase transitions (due to the lack of quark degrees of freedom). 
So to investigate the impact of QCD phase transitions on axion properties in dense medium, one must resort 
to other methods, such as the low energy effective theories and models. Among them, Nambu-Jona-Lasinio (NJL) model \cite{Nambu:1961tp,Nambu:1961fr} 
is an extensively used formalism for the study of QCD phase diagram \cite{Klevansky:1992qe,Hatsuda:1994pi,Buballa:2003qv}. 
Recently, this model has been adopted to study the low energy properties of axion in the hot and dense 
quark matter without considering the possible Cooper pairings \cite{Lu:2018ukl,Bandyopadhyay:2019pml,Abhishek:2020pjg,Zhang:2023lij}. 
 
Note that the color superconducting quark matter \cite{Alford:1997zt,Rapp:1997zu,Alford:1998mk,Rapp:1999qa,Alford:2007xm} 
may appear in the cores of compact stars. Thus it is interesting to explore how the axion is influenced by 
the diquark condensates in phases with color superconductivity (CS). Recent studies on this topic can be 
found in \cite{Balkin:2020dsr,Murgana:2024djt}. For asymptotically large baryon density, axion properties 
in the color flavor locking (CFL) phase \cite{Alford:1998mk} was calculated by employing a chiral effective 
theory in \cite{Balkin:2020dsr}. For moderate baryon number density, the coupling of the axion to two flavor color 
superconductivity (2CS) was investigated within a NJL-type model in \cite{Murgana:2024djt}, where both scalar 
and pseudo-scalar diquark condensates are considered via the instanton induced interactions. However, the 
chiral condensate and it's pseudo-scalar partner, which may play important roles near the phase boundary 
between the chiral symmetry breaking and CS phases, are all missed in \cite{Balkin:2020dsr,Murgana:2024djt}. 

In the region of low temperature T and intermediate quark chemical potential $\mu$, there may exist competition 
between the chiral and diquark condensates \cite{Alford:2007xm}. Especially, the nontrivial interplay between 
these two types of condensates may weaken the chiral transition and even lead to multiple critical end points 
\cite{Hatsuda:2006ps,Zhang:2008ima,Zhang:2009mk,Zhang:2015fda}. In order to obtain a complete insight 
on the in-medium properties of axion from the whole picture of the $T$-$\mu$ phase diagram of QCD, one must 
take into account the condensates related to Dirac-type masses in the presence of CS. The motivation of this 
work is to study the low energy properties of axion in dense medium by simultaneously including the couplings 
among the axion, quark-antiquark condensates, and diquark condensates. To do this, we adopt the two flavor NJL 
model with the one-gluon and singel-instanton exchange interactions in both the meson-meson and diquark-diquark 
channels. We mainly focus on the axion potential, axion mass, axion self coupling, and axion domain wall 
structure at moderate baryon density.

The rest of this paper is organized as follows. In section \ref{sec:model}, we describe the formalism which  
can simultaneously address the couplings between the axion field and two types of effective masses: the 
Dirac-type masses related to quark-antiquark condensates as well as the Majorana-type masses related to 
diquark condensates. In section \ref{sec:numerical}, we show numerical results and provide discussions. 
The conclusion and outlook are presented in section \ref{sec:conclusion}.

\section{The Formalism \label{sec:model}} 

In this section, we present how to simultaneously consider the couplings among the axion field and the 
scalar and pseudo-scalar condensates in both the quark-antiquark and diquark channels at the mean field 
level in the framework of two-flavor NJL model of QCD.     

\subsection{Lagrangian of a two-flavor NJL with instanton induced interactions \label{sec:LdNJL}}

There are many variants of NJL-type model \cite{Klevansky:1992qe,Hatsuda:1994pi,Buballa:2003qv}. In this 
study, we adopt a formalism of two flavor NJL model with two types of local four-quark interactions. 

The first is the color current interaction arisings from the one-gluon exchange, which takes the form 
\begin{eqnarray}
\mathcal{L}_\mathrm{int1} &=& -g_1(\bar{q}\gamma_{\mu} \lambda_a q)^2,
\label{eq:Lint1}
\end{eqnarray}
where $\lambda_a$ denote the Gell-Mann matrixes in color space and $g_1$ is the coupling constant. This 
interaction respects the global $U(2)_V\otimes U(2)_A$ symmetry in the two flavor case. The Fierz transformation
of $\mathcal{L}_\mathrm{int1}$ can give rise to different interaction forms in both the quark-antiquark and 
diquark channels (see Ref.\cite{Buballa:2003qv} for details). Here, we only consider the scalar 
and pseudo-scalar quark-antiquark channels and diquark channels, namely
\begin{eqnarray}
\mathcal{L}_\mathrm{q\bar{q}1} &=& 
G_1\left[(\bar{q}q)^2+(\bar{q}i\gamma_5{q})^2+
(\bar{q}\vec\tau{q})^2+
(\bar{q}i\gamma_5\vec\tau q)^2\right]
\label{eq:Lint1qaq}
\end{eqnarray}
with $G_1=\frac{N_c^2-1}{N_c^2}g_1$ and 
\begin{eqnarray}
\mathcal{L}_\mathrm{qq1} &=& 
 H_1\sum_{A}\big[ (q^T C i\gamma_5 
 \tau_2\lambda_A q) 
 (\bar q i \gamma_5 C\tau_2\lambda_A\bar q^T)
\nonumber\\&&  
+ (q^T C \tau_2\lambda_A q) 
 (\bar q  C   \tau_2\lambda_A\bar  q^T)
 \big]
\label{eq:Lint1qq}
\end{eqnarray}
with $H_1=\frac{N_c+1}{2N_c}g_1$, where $N_c=3$ is the color number and $\lambda_A$ (A=2,5,7) are the antisymmetric 
Gell-Mann matrixes in color space. The $\tau_a$ in Eqs.\eqref{eq:Lint1qaq} and \eqref{eq:Lint1qq} are 
Pauli matrixes in flavor space and $C=i\gamma^2\gamma^0$  is the charge conjugate matrix. Note that only 
the flavor and color antisymmetric diquark channels are listed in Eq.\eqref{eq:Lint1qq}. We see that the 
standard ratio $H_1/G_1$ in the Fierz transformation is $3/4$ for $N_c=3$.  

Another one is the singe instanton induced four-quark interaction for the two flavor case, which reads
\begin{eqnarray}
\mathcal{L}_\mathrm{int2}&&= 
\frac{g_2}{4(N_c^2-1)}
\bigg\{\frac{2N_c-1}{2N_c}\times
\nonumber\\&&
\left[(\bar{q}q)^2-(\bar{q}i\gamma_5{q})^2-
(\bar{q}\tau_a{q})^2+
(\bar q \tau_a i\gamma_5 q)^2\right]
\nonumber\\
&& -\frac{1}{4N_c}\left[(\bar{q}\sigma^{\mu\nu}q)^2-(\bar{q}\sigma^{\mu\nu}\tau_a{q})^2 \right] 
\bigg\}
\label{eq:Lint2}
\end{eqnarray}
according to Ref.\cite{Rapp:1999qa}. 
This interaction respects the global $U(1)_V\otimes {SU(2)_V} \otimes SU(2)_A$ symmetry but 
violates the global $U(1)_A$ symmetry explicitly, which is usually used to describe the axial  
anomaly of QCD in the NJL-type models. Performing the Feirz transformation of $\mathcal{L}_\mathrm{int2}$,  
we obtain the scalar and pseudo-scalar quark-antiquark interactions        
\begin{eqnarray}
\mathcal{L}_\mathrm{q\bar{q}2} &=& 
G_2\left[(\bar{q}q)^2-(\bar{q}i\gamma_5{q})^2
-(\bar{q}\vec\tau{q})^2+(\bar{q}i\gamma_5\vec\tau q)^2\right]
\label{eq:Lint2qaq}
\end{eqnarray}
and the scalar and pseudo-scalar diquark interactions 
\begin{eqnarray}
\mathcal{L}_\mathrm{qq2} &=&
 H_2\sum_{A}
 \big[ (q^T C i\gamma_5 
 \tau_2\lambda_A q) 
 (\bar q i \gamma_5 C\tau_2\lambda_A\bar q^T)
\nonumber\\&&  
 - (q^T C \tau_2\lambda_A q) 
 (\bar q  C   \tau_2\lambda_A\bar  q^T)
 \big],
\label{eq:Lint2qq}
\end{eqnarray}
where $G_2=\frac{g_2}{4N_c^2}$ and $H_2=\frac{ g_2}{8N_c(N_c-1)}$ \cite{Rapp:1999qa,Buballa:2003qv}. 
Similar to $\mathcal{L}_\mathrm{qq1}$, the color and flavor symmetric diquark channels are ignored 
in $\mathcal{L}_\mathrm{qq2}$. The Feirz transformation also gives rise to the ratio $H_2/G_2=3/4$ 
for $N_c=3$.   

As mentioned, we will simultaneously take into account the scalar and pseudo-scalar condensates 
in both the quark-antiquark and diquark channels. We adopt the following Lagrangian density 
\begin{equation}
\mathcal{L} = 
\bar q \left(
i \rlap/\partial
+ \hat\mu\gamma_0 - m_0
\right) q 
+\mathcal{L}_\mathrm{q\bar{q}}
+\mathcal{L}_\mathrm{qq},
\label{eq:Lgeff}
\end{equation}
where
\begin{eqnarray}
\mathcal{L}_\mathrm{q\bar{q}} =
 \mathcal{L}_\mathrm{q\bar{q}1}+\mathcal{L}_\mathrm{q\bar{q}2},
 \label{eq:intqaq}
\end{eqnarray}
and
\begin{eqnarray}
\mathcal{L}_\mathrm{qq} =
\mathcal{L}_\mathrm{qq1}+\mathcal{L}_\mathrm{qq2}.
 \label{eq:intqq}
\end{eqnarray}
Note that in Ref.\cite{Murgana:2024djt}, only the diquark interactions \eqref{eq:intqq} are   
taken into account and the quark masses are ignored for simplicity.  

We use the coupling constant $G$ and a dimensionless parameter $c$ to indicate the couplings $G_1$ 
and $G_2$ through the relations
\begin{eqnarray}
 G_2=c G, \quad G_1=(1-c)G.
\label{eq:Gc}
\end{eqnarray}
The parameter $G$ can be fixed by the vacuum properties of QCD. 

\subsection{ Lagrangian of NJL with axion field \label{sec:NJLAxion}}

To introduce the axial filed conveniently, it is more useful to express the Lagrangian density \eqref{eq:Lgeff} 
in term of the left(right)-handed quark field $q_{L(R)}=\mathcal{P}_{L(R)} {q}$, where
\begin{equation}
\mathcal{P}_R = \frac{1+\gamma_5}{2},~~~
\mathcal{P}_L = \frac{1-\gamma_5}{2}.
\label{eq:projectors}
\end{equation}   
Using $q_{L(R)}$, the interaction $\mathcal{L}_\mathrm{q\bar{q}2}$ can be rewritten 
as the sum of two determinants   
\begin{eqnarray}
\mathcal{L}_\mathrm{q\bar{q}2} = 
8G_2\left[
\mathrm{det}(\bar q_R q_L) + \mathrm{det}(\bar q_L q_R) 
\right],
\label{eq:Lintdet}
\end{eqnarray}
and the diquark interaction $\mathcal{L}_\mathrm{q{q}2}$ takes the form  
\begin{eqnarray}
\mathcal{L}_\mathrm{q{q}2} = 
-2H_2\sum_{A}\left[
d_{A,R}^{\dag} d_{A,L} +
d_{A,L}^{\dag} d_{A,R} 
\right],
\label{eq:Lintqq22}
\end{eqnarray}
where
\begin{eqnarray}
d_{A,L(R)}=q^T_{L(R)} C i \tau_2 \lambda_A q_{L(R)}. 
\label{eq:qqLR}
\end{eqnarray} 

Performing the following $U(1)_A$ transformation  
\begin{eqnarray}
q_{L} \rightarrow e^{-i\alpha}q_{L} \, \, \, \text{and} \, \, \, q_{R} \rightarrow e^{i\alpha}q_{R}, 
\label{eq:qqLR}
\end{eqnarray}
the interactions $\mathcal{L}_\mathrm{q\bar{q}2}$ and $\mathcal{L}_\mathrm{q{q}2}$ become
\begin{eqnarray}
\mathcal{L}_\mathrm{q\bar{q}2}\rightarrow 
8G_2\left[
e^{-i4\alpha}\mathrm{det}(\bar q_R q_L) +
e^{i4\alpha}\mathrm{det}(\bar q_L q_R) 
\right]
\label{eq:detUA}
\end{eqnarray}
and
\begin{eqnarray}
\mathcal{L}_\mathrm{q{q}2}\rightarrow
-2H_2\sum_{A}\left[
e^{-i4\alpha}d_{A,R}^{\dag} d_{A,L} +
e^{i4\alpha}d_{A,L}^{\dag} d_{A,R} 
\right],\nonumber\\
\label{eq:dqUA}
\end{eqnarray} 
respectively. The axion field is then introduced by replacing the phase factor $4\alpha$ (namely $2N_f\alpha$ 
for $N_f=2$) in Eqs.\eqref{eq:detUA} and \eqref{eq:dqUA} with $a/f_a$, where $a$ is the axion field. So the NJL 
lagrangian density with the axion field reads
\begin{eqnarray}
\mathcal{L}_{\text{eff}} &=& 
\bar q \left(
i \rlap/\partial
+ \hat\mu\gamma_0 - m_0
\right) q \nonumber\\
&&+\mathcal{L}_\mathrm{q\bar{q}1}
+\mathcal{L}_\mathrm{qq1}+\mathcal{L}_\mathrm{aq\bar{q}}+\mathcal{L}_\mathrm{aqq},
\label{eq:NJLaxion}
\end{eqnarray}
where
\begin{eqnarray} 
\mathcal{L}_\mathrm{aq\bar{q}}=8G_2\left[
e^{-i\frac{a}{f_a}}\mathrm{det}(\bar q_R q_L) +
e^{i\frac{a}{f_a}}\mathrm{det}(\bar q_L q_R)\right]
\end{eqnarray} 
and
\begin{eqnarray} 
\mathcal{L}_\mathrm{aqq}&=&-2H_2\sum_{A}\left[
e^{-i\frac{a}{f_a}}d_{A,R}^{\dag} d_{A,L} +
e^{i\frac{a}{f_a}}d_{A,L}^{\dag} d_{A,R} 
\right].
\label{eq:int2qqform1} 
\end{eqnarray} 
We can rewrite the interaction $\mathcal{L}_\mathrm{aqq}$  as   
\begin{eqnarray}
\mathcal{L}_\mathrm{aqq} &=&
 H_2\sum_{A}
 \big[ (\bar q i \gamma_5 C\tau_2\lambda_A\bar q^T)(q^T C i\gamma_5 
 \tau_2\lambda_A q)\cos(\frac{a}{f_a})
\nonumber\\&&
-(\bar q i \gamma_5 C\tau_2\lambda_A\bar q^T)(q^T C \tau_2\lambda_A q)\sin(\frac{a}{f_a})
\nonumber\\&&
-(\bar q C\tau_2\lambda_A\bar q^T)(q^T C i\gamma_5 
 \tau_2\lambda_A q)\sin(\frac{a}{f_a})  
\nonumber\\&&  
+(\bar q  C i \tau_2\lambda_A\bar  q^T)(q^T C i \tau_2\lambda_A q)\cos(\frac{a}{f_a})
 \big].
\label{eq:int2qqform2} 
\end{eqnarray} 

Note that the diquark interaction $\mathcal{L}_\mathrm{q{q}1}$ can also be written  
in term of $d_{A,R}$ and $d_{A,L}$, which reads  
\begin{eqnarray}
\mathcal{L}_\mathrm{q{q}1} = 
2H_1\sum_{A}\left[
d_{A,R}^{\dag} d_{A,R} +
d_{A,L}^{\dag} d_{A,L} 
\right].
\label{eq:Lintqq12}
\end{eqnarray}
Clearly this interaction is $U(1)_A$-preserving and thus doesn't couple with the axion field.

\subsection{ Nambu-Gorkov propagator with quark-antiquark and diquark condensates \label{sec:NG}}

For nonzero $\theta=a/f_a$, four condensates, namely the chiral condensate $\sigma$, the speudo-scalar quark 
condensate $\eta$, the scalar diquark condensate $\delta$, and the speudo-scalar diquark condensate $\omega$,  
may appear at low temperature and moderate density. These condensates are defined as 
\begin{eqnarray}
&&\langle
\bar{q} q \rangle  = \sigma, \label{eq:sigma}\\
&&\langle
\bar{q}  i\gamma_5 q \rangle = \eta, \label{eq:yita}\\
&&\langle
q^T C  i\gamma_5 \tau_2\lambda_2 q \rangle  = \delta, \label{eq:s2cs}\\
&&\langle
\bar{q} i\gamma_5 \tau_2\lambda_2  C \bar{q}^T \rangle  = \delta^{*}, \label{eq:cs2cs}\\
&&\langle
q^T C i \tau_2\lambda_2 q \rangle = \omega, \label{eq:ps2cs}\\
&&\langle
\bar{q} i \tau_2\lambda_2 C \bar{q}^T \rangle   = -\omega^{*}. \label{eq:cps2cs}
\end{eqnarray}
Following the convention, red and green quarks are assumed to participate the Cooper pairing 
and thus only $\lambda_2$ appears in Eqs.\eqref{eq:s2cs}-\eqref{eq:cps2cs}. We can also define the 
following left and right handed diquark condensates
\begin{eqnarray}
&&\langle
d_{2,L} \rangle  = \langle q^T_L C i \tau_2 \lambda_2 q_L\rangle = h_L, \label{eq:L2cs}\\
&&\langle
d_{2,L}^{\dag} \rangle  = -\langle \bar{q}_L C i \tau_2 \lambda_2 \bar{q}_L^{T}\rangle = h_L^{*}, \label{eq:Lc2cs}\\
&&\langle
d_{2,R} \rangle  =\langle q^T_R C i \tau_2 \lambda_2 q_R\rangle = h_R. \label{eq:R2cs}\\
&&\langle
d_{2,R}^{\dag} \rangle  = -\langle \bar{q}_R C i \tau_2 \lambda_2 \bar{q}_R^{T}\rangle = h_R^{*}. \label{eq:Rc2cs}
\end{eqnarray} 
Clearly, there exists the relations
\begin{eqnarray}
\delta=h_R-h_L,\,\,\, \omega=h_L+h_R\nonumber\\
\delta^{*}=h_R^{*}-h_L^{*},\,\,\, \omega^{*}=h_L^{*}+h_R^{*}. \label{eq:swhh}
\end{eqnarray}
Note that $h_{L(R)}$ introduced above corresponds to $-h_{L(R)}$ used in \cite{Murgana:2024djt}. Since both $\tau_2$ 
and $\lambda_2$ in Eqs.\eqref{eq:L2cs}-\eqref{eq:Rc2cs} are imaginary matrixes, the relation \eqref{eq:swhh}   
is still the same as that in \cite{Murgana:2024djt}.   
         
We will adopt the mean field treatment in this paper. Using the assumptions $\delta=\delta^{*}$ and $\omega=\omega^{*}$ 
(or $h_L=h_L^{*}$ and $h_R=h_R^{*}$ as that in \cite{Murgana:2024djt}.),  we get the following mean field Lagrangian 
related to the couplings $H_1$ and $H_2$ 
\begin{eqnarray}
\mathcal{L}_\mathrm{Mqq} &=&
 H_1 
\left[\delta(\bar q i \gamma_5 C \tau_2\lambda_2\bar q^T) 
+\delta^{*}(q^T C i\gamma_5 \tau_2\lambda_2 q) \right]
\nonumber \\
&&- H_1 
\left[\omega(\bar q i  C\tau_2\lambda_2 \bar q^T) 
-\omega^{*}(q^Ti C \tau_2\lambda_2 q) \right]
\nonumber \\
&&+ H_2  
\left[
\delta(\bar q  i\gamma_5 C\tau_2\lambda_2
\bar  q^T) +\delta^{*}
(q^T C i\gamma_5 \tau_2\lambda_2 q) 
\right]\cos\frac{a}{f_a}\nonumber\\
&&
+H_2\left[-\delta^{*}(q^T C \tau_2\lambda_A q)-\omega(\bar q \gamma_5 C\tau_2\lambda_A\bar q^T)\right]\sin\frac{a}{f_a}
\nonumber\\
&&
+H_2\left[-\delta(\bar q C\tau_2\lambda_A\bar q^T) +\omega^{*}(q^T C \gamma_5 
 \tau_2\lambda_A q) \right]\sin\frac{a}{f_a}
\nonumber\\&&  
+H_2\left[\omega(\bar q  C i \tau_2\lambda_A\bar  q^T)-\omega^{*}(q^T C i \tau_2\lambda_A q)\right]\cos\frac{a}{f_a}
\nonumber\\&& 
- H_1\left[ |\delta|^2+|\omega|^2\right]-H_2 \left[|\delta|^2-|\omega|^2 \right]\cos\frac{a}{f_a}.
\label{eq:Lmwitha}
\end{eqnarray}
The mean field interaction Lagrangian related to $G_1$ and $G_2$ is the same as that in \cite{Lu:2018ukl}.  
  
To study the quark Cooper pairings, it is more convenient to use the Nambu-Gorkov formalism \cite{Buballa:2003qv}. 
So we introduce the following bi-spinors of quark fields  
\begin{equation}
     \Psi=\left(\begin{array}{c}
           \frac{1}{\sqrt{2}} q\\
          \frac{1}{\sqrt{2}} C \bar q^T
     \end{array}\right), \qquad \bar \Psi= \left(\frac{1}{\sqrt{2}}\bar q, \frac{1}{\sqrt{2}}q^T C\right).
\label{eq:NGF}     
 \end{equation}
The interaction Lagrangian density with the axial field at the mean field level is then rewritten as   
\begin{equation}
    \mathcal{L} = \bar\Psi S^{-1} \Psi-\mathcal{V},
\end{equation}   
where $S^{-1}$ is the inverse Nambu-Gorkov quark propagator and
\begin{eqnarray}
    \mathcal{V}&=&{G_1}\left( {{\eta ^2} + {\sigma ^2}} \right) - {G_2}\left( {{\eta ^2}-{\sigma^2}} \right)\cos \frac{a}{f_a} + 2{G_2}\sigma \eta \sin \frac{a}{f_a}\nonumber
    \\
    &&+H_1( \delta^2+\omega^2)+H_2 (\delta^2-\omega^2)\cos\frac{a}{f_a}. \label{eq:VacEenergy} 
\end{eqnarray} 
In the presence of the condensates $\sigma$, $\eta$, $\delta$, and $\omega$, the matrix $S^{-1}$ in the 
momentum space takes the form   
\begin{equation}
 S^{-1}(p)=\left(
 \begin{array}{cc}
(\slashed{p}_{+}-M)\bm 1_C\bm 1_F 
\ &  \Phi^-  \\
         &\\
       \Phi^+  &  (\slashed{p}_{-}-M)\bm 1_C\bm 1_F 
     \end{array}\right),
     \label{eq:NGP}
\end{equation}  
where
\begin{eqnarray}
\slashed{p}_{\pm}&=&\slashed{p}\pm \mu\gamma_0,
\\
M&=&M_s-i\gamma_5M_p, \label{eq:Mass}
\\
\Phi^- &=& \Delta_s[i\gamma_5\tau_2\lambda_2]+\Delta_p\tau_2\lambda_2,\label{eq:S12}
    \\
\Phi^+ &=& \Delta_s^*[i\gamma_5\tau_2\lambda_2]+\Delta_p^*\tau_2\lambda_2, \label{eq:S21}
\end{eqnarray}
and $\bm 1_{C(F)}$ is the identity matrix in color (flavor) space. 

Four energy gaps appear in Eqs.\eqref{eq:Mass}-\eqref{eq:S21}, namely the scalar (psudo-scalar)
Dirac-type mass $M_s$($M_p$) and the scalar (psudo-scalar) Majorana-type mass $\Delta_s$($\Delta_p$), which 
are defined as  
\begin{eqnarray}
  M_s& =& {m_0}-2\left( {{G_1} + {G_2}\cos \frac{a}{{{f_a}}}} \right)\sigma - 2{G_2}\eta \sin \frac{a}{f_a}, \label{eq:SMass}
  \\
  M_p& =& -2\left( {{G_1} - {G_2}\cos \frac{a}{{{f_a}}}} \right)\eta  - 2{G_2}\sigma \sin \frac{a}{f_a}, \label{eq:PMass}
  \\
  \Delta_s &=& (2{H_1} + 2{H_2}\cos\frac{a}{f_a} )\delta + i2{H_2}\omega\sin\frac{a}{f_a}, \label{eq:SMjMass}
  \\
  \Delta_p &=& -2{H_2}\delta\sin \frac{a}{f_a} - i(2{H_1} - 2{H_2}\cos \frac{a}{f_a} )\omega. \label{eq:PMjMass}
\end{eqnarray}
Unlike $M_s$ and $M_p$, we see that $\Delta_s$ and $\Delta_p$ are complex quantities for nonzero axion field. 

\subsection{Thermodynamic potential\label{sec:ld}}

As in the previous studies \cite{Lu:2018ukl,Murgana:2024djt}, the axion field is treated as a classical 
background here. Performing the standard functional integration over the quark fields, we can obtain the mean field 
thermodynamical potential at a fixed $a/f_a$, which reads  
\begin{equation}
    \Omega=\mathcal{V}+\Omega_{q},
\end{equation}
where
\begin{equation}
  \Omega_{q}=-T\sum_n\int \frac{d^3 \,p}{(2\pi)^3} \frac{1}{2}
  \mathrm{Tr}
  \ln  \left(\frac{1}{T} S^{-1}(i\omega_n, \vec{p})\right) 
  \label{eq:one-loop}
\end{equation}
is the one-loop contribution of the fermions. In \eqref{eq:one-loop}, $w_n=(2n+1)\pi T$ are the Matsubara 
frequencies for fermions in the imaginary time thermal field theory and the trace is taken over the 
Nambu-Gorkov, Dirac, color, and flavor spaces. The added overall factor $1/2$ is used to cancel the doubling 
of degrees of freedom due to the use of bi-spinors \cite{Buballa:2003qv}.      

The Matsubara summation in \eqref{eq:one-loop} can be simplified significantly once we obtain 
the eigenvalues of the following matrix (for more details on the method see the appendix of 
Ref.~\cite{Ruester:2005jc} or \cite{Blaschke:2005uj}).      
\begin{eqnarray}    
 \mathcal{Z}&=&\gamma^{0}S^{-1}(p_0,\vec{p})-p_0\bm 1\\
 &=&\left(
 \begin{array}{cc}
-\vec{\gamma}\cdot \vec{p}+\mu-\gamma^{0}M \ &  \gamma^{0}\Phi^- \\
         &\\
       \gamma^{0}\Phi^+  &  -\vec{\gamma}\cdot \vec{p}-\mu-\gamma^{0}M
     \end{array}\right).
\end{eqnarray} 
The traceless property of this matrix indicates that if the value $E_i$ is one of its eigenvalues then $-E_i$ may 
also be, which has been confirmed in our calculation. Even the matrix $\mathcal{Z}$ is more complicated than the 
corresponding one without the mass $M$ in \cite{Murgana:2024djt}, we can still get the analytical eigenvalues 
which read
\begin{eqnarray}
  {E_{1,\pm}} &=& \pm \sqrt{{(E - \mu)}^2}=\pm \epsilon_1(\vec{p}), \label{eq:egv1}\\
  {E_{2,\pm}} &=& \pm \sqrt{{(E + \mu)}^2}=\pm \epsilon_2(\vec{p}), \label{eq:egv2}\\
  {E_{3,\pm}} &=& \pm\sqrt {{E'^2} + {\mu ^2} - Z_ + ^2}=\pm \epsilon_3(\vec{p}),  \label{eq:egv3} \\
  {E_{4,\pm}} &=& \pm\sqrt {{E'^2} + {\mu ^2} + Z_ + ^2}=\pm \epsilon_4(\vec{p}),  \label{eq:egv4}  \\
  {E_{5,\pm}} &=& \pm\sqrt {{E'^2} + {\mu ^2} - Z_ - ^2}=\pm \epsilon_5(\vec{p}),  \label{eq:egv5} \\
  {E_{6,\pm}} &=& \pm\sqrt {{E'^2} + {\mu ^2} + Z_ - ^2}=\pm \epsilon_6(\vec{p}),  \label{eq:egv6}
\end{eqnarray}
where
\begin{eqnarray}
{E} &=& \sqrt {{p^2} +  M_s^2+M_p^2}\\
E'^2 &=& p^2 +  M_s^2+M_p^2+ |\Delta_s|^2 + |\Delta_p|^2
\end{eqnarray}   
and $p=|\vec{p}|$. The terms $Z_\pm^2$ reflect the mixing between the Dirac and 
Majorana masses which are defined as 
\begin{eqnarray}
\begin{split}
   Z_ \pm ^2 & =\big\{\left[|\Delta_s^*\Delta_p-\Delta_s{\Delta_p^*}| \pm 2p\mu \right]^2 + 4\big[{\left| \Delta_s \right|^2}M_p^2 \\
    &+ {\left| \Delta_p \right|^2}M_s^2 + {M^2}{\mu ^2} - {M_s}{M_p}({\Delta_s^*}\Delta_p + \Delta_s{\Delta_p^*})\big]\big\}^{1/2},
\end{split}
\label{eq:Mixing}
\end{eqnarray}
where
\begin{eqnarray}
|\Delta_s^*\Delta_p-\Delta_s{\Delta_p^*}|=8(H_1^2 - H_2^2)\delta\omega.
\end{eqnarray}
Since the inverse Nambu-Gokov propagator is a $48 \times 48$ matrix, only twelve independent eigenvalues 
displayed in \eqref{eq:egv1}-\eqref{eq:egv6} imply that the multiplicity of each eigenvalue equals 
to four.  

We see that eigenvalues $E_{1,\pm}$ and ${E_{2,\pm}}$ only depend on the Dirac masses but the 
ones from $E_{3,\pm}$ to $ E_{6,\pm}$ depend both on the Dirac and Majorana masses. Clearly, the 
former corresponds to the dispersion relations of the blue quarks and the later the red and green ones. 
We can check that for vanishing $M_s$ and $M_p$, Eqs.\eqref{eq:egv1}-\eqref{eq:egv6} can be reduced 
to the corresponding dispersion relations obtained in \cite{Murgana:2024djt} where only the Majorana 
masses are considered. On the other hand, the standard quark dispersion relations for 2CS \cite{Alford:2007xm} 
can be reproduced from Eqs.\eqref{eq:egv1}-\eqref{eq:egv6} by fixing $a/f_a=0$ (and thus $M_p=\Delta_p=0$). 
 
Using the identity 
\begin{equation}
  \mbox{Tr}\ln  \frac{1}{T} S^{-1}=\ln\det \frac{1}{T} S^{-1},\nonumber
\end{equation}
we can decompose the trace in \eqref{eq:one-loop} as  
\begin{equation}
     \mbox{Tr}\ln \frac{1}{T} S^{-1}(i\omega_n, \vec{p})= 4\sum_{k=1}^{6} \ln \left(\frac{\omega_n^2+\epsilon_k(\vec{p})^2}{T^2}\right), \label{eq:trace}
\end{equation}
where $\epsilon_k$ are the six independent positive eigenvalues listed in Eqs.\eqref{eq:egv1}-\eqref{eq:egv6} 
and the factor four in the RHS of \eqref{eq:trace} is the eigenvalue degeneracy.  
Now the Matsubara summation \eqref{eq:one-loop} can be evaluated analytically by employing the standard relation \cite{Kapusta:2023eix}
\begin{equation}
\sum_n \ln\left(\frac{\omega_n^2+\varepsilon^2}{T^2}\right) =\frac{|\varepsilon|}{T}+ 2 \ln \left(1+e^{-|\varepsilon|/T}\right).
\end{equation}
 We then obtain the following mean field thermodynamic potential
\begin{eqnarray}
\Omega &=& -2\sum_{k=1}^{6}\int \frac{d^3p}{(2\pi)^3} \left[\epsilon_k(\vec{p})+ 2T \ln \left(1+e^{-\epsilon_k(\vec{p})/T}\right)\right]\nonumber\\
&&+\mathcal{V}.
\label{eq:MeanTP}  
\end{eqnarray}

For a given value of $a/f_a$ at fixed $T$ and $\mu$, the thermodynamical potential is a function of 
$\sigma$, $\eta$, $\delta$, and $\omega$. The physical values of these condensates are determined by the 
following gap equations
\begin{eqnarray}
\frac{\partial{\Omega}}{\partial{\sigma}}=0, \,\, \frac{\partial{\Omega}}{\partial{\eta}}=0, \,\, \frac{\partial{\Omega}}{\partial{\delta}}=0, \,\,\frac{\partial{\Omega}}{\partial{\omega}}=0. \label{eq:GapEqs}
\end{eqnarray}   

\subsection{Axion potential, axion mass, and self-coupling \label{sec:Amass}}

The effective axion potential can be defined as
\begin{equation}
V(a,T,\mu)=\Omega(x_i(a,T,\mu),a,T,\mu), \label{eq:AP}
\end{equation} 
where $x_i(a,T,\mu)$ refer to the physical values of the aforementioned four condensates 
obtained at given $a$, $T$, and $\mu$.

By taking the second derivative of the potential \eqref{eq:AP} with respect to $a$ 
at $a=0$, we can obtain the axion mass squared 
\begin{eqnarray}
m_a^2=\left.\frac{d^2 {V} }{da^2}\right|_{a=0} =\frac{\chi_{t}}{f_a^2}, \label{eq:amass}
\end{eqnarray}
where $\chi_{t}$ is the topological susceptibility. 
The axion quartic self-coupling is defined as the fourth derivative of \eqref{eq:AP} at $a=0$, which reads  
\begin{eqnarray}
\lambda _a=\left.\frac{d^4{V} }{da^4}\right|_{a=0}. \label{eq:acoupling}
\end{eqnarray}

Since the physical condensates are all implicitly dependent on $a$, the total differential of $V(a)$ 
with respect to $a$ satisfies the following relation
\begin{equation}
 \frac{\mathrm{d}{V}}{\mathrm{d} a}=\frac{\partial {V}}{\partial a}+\frac{\partial {V}}{\partial \sigma }\frac{\partial \sigma }{\partial a}
 +\frac{\partial {V}}{\partial \eta}\frac{\partial \eta }{\partial a} +\frac{\partial {V}}{\partial \delta}\frac{\partial \delta }{\partial a} 
 +\frac{\partial {V}}{\partial \omega}\frac{\partial \omega }{\partial a}.
\end{equation}
Therefore to evaluate the axion mass and self-coupling, we need to calculate the 1-4th partial derivatives 
of each of the physical condensates with respect to $a$. This can be fulfilled by taking the successive derivatives 
of the gap equations \eqref{eq:GapEqs} with respect to $a$.    

\section{\label{sec:numerical} NUMERICAL RESULTS AND DISCUSSIONS}

In this section, we present the numerical results obtained in the NJL model. We focus on the axion 
potential, axion mass, and axion self coupling at finite $T$ and $\mu$. The properties of axion domain 
walls in the presence of 2CS are also reported. 

We adopt the same model parameters as that in \cite{Lu:2018ukl}, namely $\Lambda=590\,\text{MeV}$, 
$G=2.435/\Lambda^2$, $c=0.2$, and $m=6\, \text{MeV}$. These parameters are fixed by fitting the physical 
pion mass, the pion decay constant, and the chiral condensate $\sigma_0=2 (-241.5 \,\text{MeV})^3$ in 
vacuum. The ratios $H_1/G_1$ and $H_2/G_2$ are fixed by the Fierz transformations. 

Note that the sensitivities of some results to the variation of $c$ in the range $c=(0,0.5)$ are also investigated. 
In addition, the axion mass and self coupling obtained using the ratios $H_1/G_1$ and $H_2/G_2$ beyond the 
Fierz transformation are given in Appendix.      

In the following, the unit $\text{MeV}$ in the point $(T,\mu)$ is ignored for simplicity. 

\subsection{The condensates as functions of ${a}/{f_a}$ \label{subs: condensates} }

In this subsection, we show how the condensates $\sigma$, $\eta$, $\delta$, and $\omega$ vary with  
the angle $\theta \equiv {a/f_a}$ at finite $T$ and $\mu$. 

\begin{figure}[ht]
\begin{center}
   \includegraphics[scale=0.35]{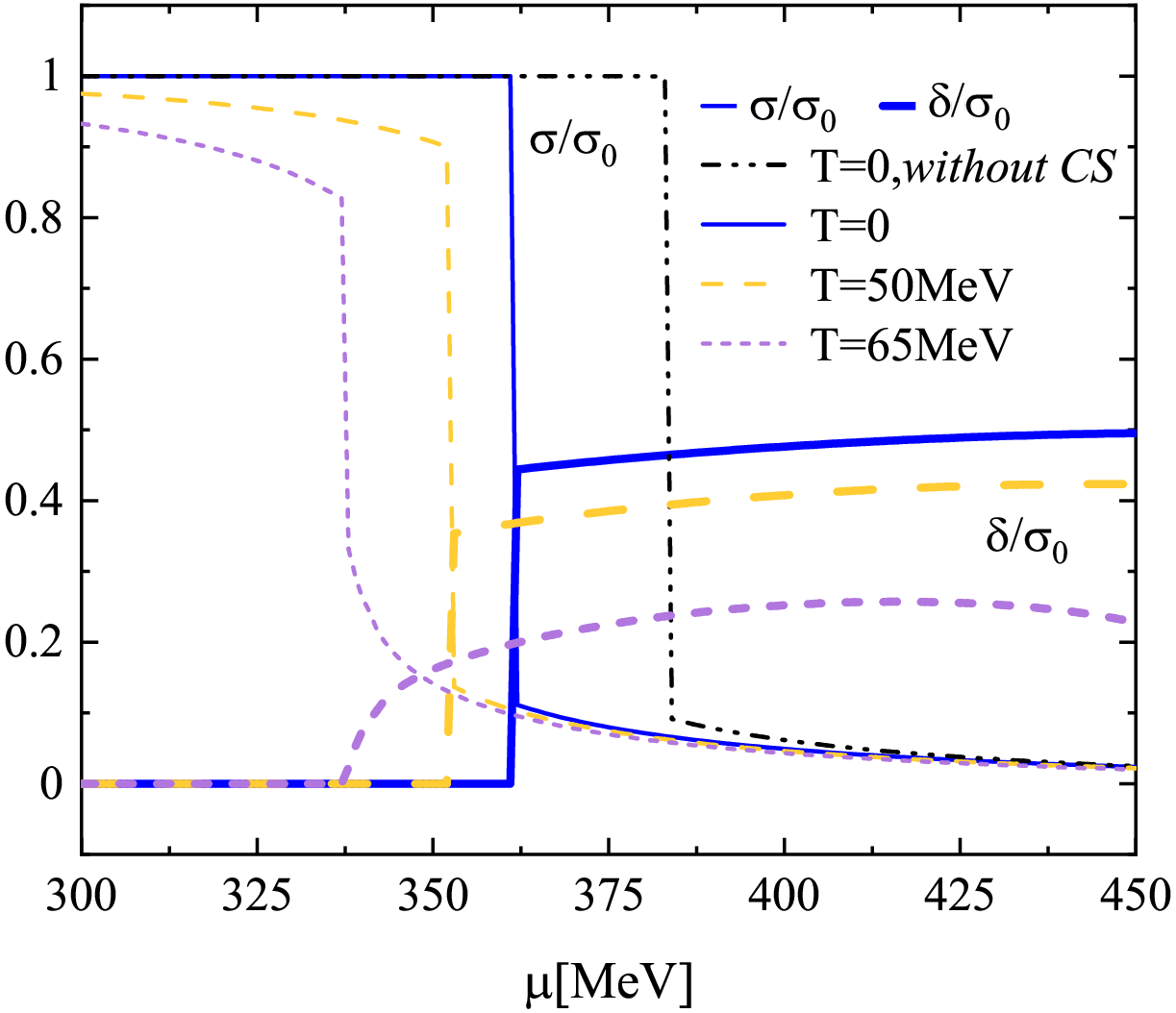}
\end{center}
\caption{The chiral condensate $\sigma$ (thin lines) and diquark condensate $\delta$ (bold lines) as functions 
of $\mu$ for several values of $T$ at ${a}/{f_a}=0$. Both condensates are scaled by the vacuum chiral condensate 
${\sigma_0}$. The black dash-doted-doted line corresponds to $\sigma $ at $T=0$ without considering the CS. }
\label{fg:cden-zeroa}
\end{figure}

In Fig.\ref{fg:cden-zeroa}, we display the chiral condensate $\sigma $ and diquark condensate $\delta$ as 
functions of $\mu$ for several values of $T$ at ${a}/{f_a}=0$. In this case, the psudo-scalar condensates 
$\eta$ and $\omega$ are both vanishing.  For $T=0$, the first order chiral transition happens at $\mu\simeq 361\,\text{MeV}$, 
at which $\sigma$ drops significantly and a finite $\delta$, which corresponds to the gap $\Delta_s = 140\, \text{MeV}$,  
appears. The critical chemical potential at $T=0$ reduces by $\sim25\, \text{MeV}$ compared to the case without 
considering the CS (see the black dash-doted-doted line). We see that the first order transition is weakened with 
rising $T$ and the 2CS still emerges at $T=65\,\text{MeV}$. 

\begin{figure}[ht]
 \begin{center}
   \includegraphics[scale=0.35]{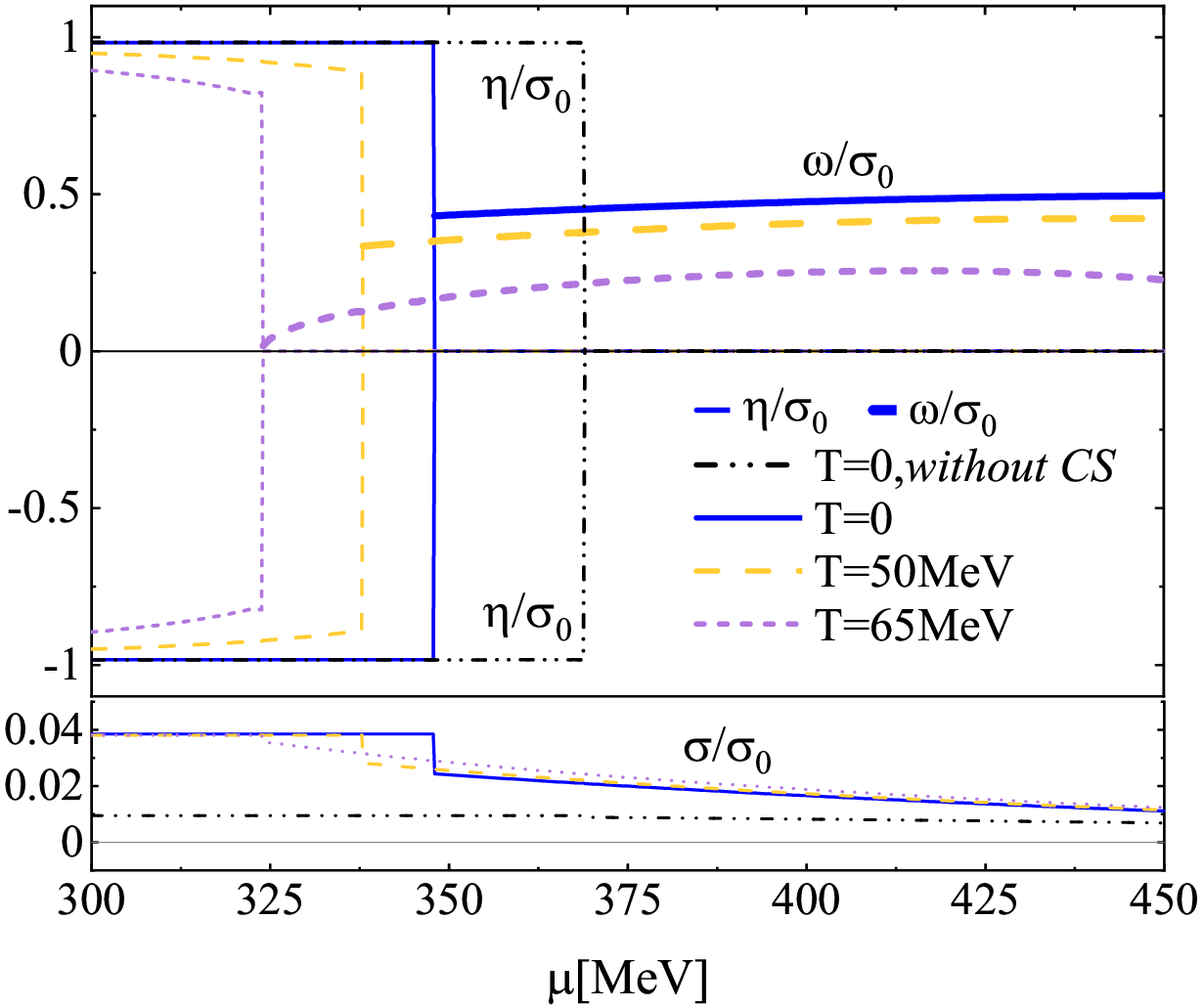}
   \end{center}
   \caption{ The condensates $\sigma$,  $\eta$, and $\omega $ as functions of $\mu$ for several 
   values of $T$ at $a/{f_a} = \pi$. In the upper panel, the thin (bold) lines refer to $\eta$ ($\omega$).  
   All the quantities are normalized by ${\sigma_0}$.}
   \label{fg:wupi}
 \end{figure}

\begin{figure}[ht]
   \begin{center}
   \includegraphics[scale=0.35]{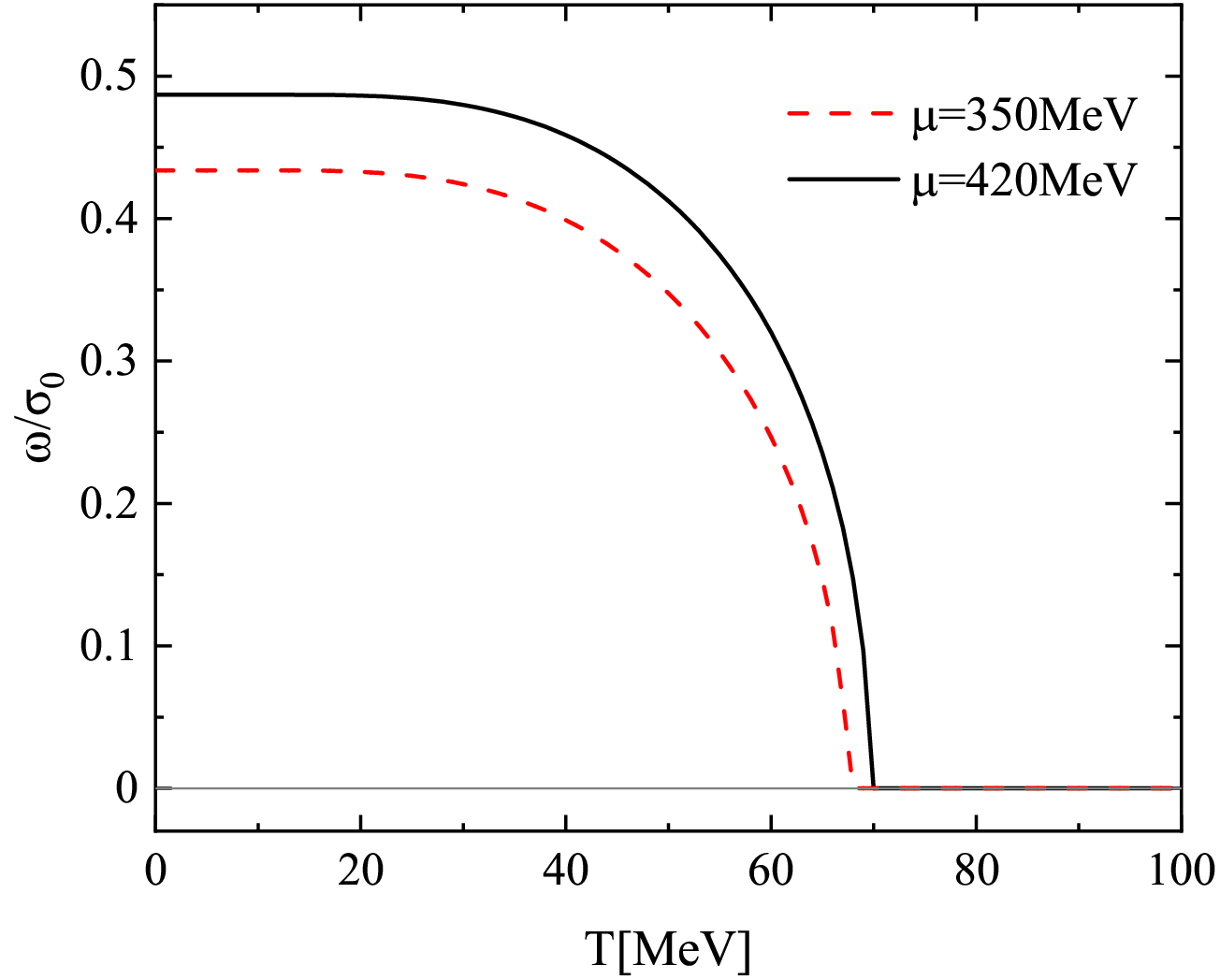}
   \end{center}
   \caption{ The normalized pseudo-scalar diquark condensate $\omega$ versus $T$ for $\mu=350$ 
   and $420\text{MeV}$ at $a/f_a=\pi$.}
   \label{wtpi}
\end{figure}

Figure \ref{fg:wupi} displays $\sigma$, $\eta$, and $\omega$ as functions of $\mu$ at $a/f_a=\pi$ for 
the same values of $T$ as that in Fig.\ref{fg:cden-zeroa}. Compared to Fig.\ref{fg:cden-zeroa}, the roles 
of the scalar condensate $\sigma$ ($\delta$) and the pseudo-scalar condensate $\eta$ ($\omega$) exchange:  
at lower quark chemical potentials, $\eta$ dominates and $\sigma$ becomes very small (but nonzero due to 
the small current quark mass); at larger quark chemical potentials, $\omega$ dominates and $\delta$ vanishes. 
This indicates the spontaneous breaking of the parity symmetry for lower temperatures at $a/f_a=\pi$. We notice 
that $\eta$ exhibits double values with the same magnitude but opposite sign for $\mu<\mu_c(T)$. This is 
just the so called Dashen's phenomena\cite{Dashen:1971} characterized by the two-fold vacuum degeneracy 
at $a/f_a=\pi$. For $\mu>\mu_c(T)$, $\eta$ becomes zero and the Dashen's phenomena breaks down, which 
is similar to the case at finite temperature where Dashen's phenomena only holds for $T<T_c$. Fig.\ref{fg:wupi}
shows that at $T=0$, the $\eta$ degeneracy is lifted for $\mu>\mu_c=347\, \text{MeV}$ where $\eta$ drops 
to zero and $\omega$ appears.  
In contrast to the case without considering the CS, the critical chemical potential $\mu_c$ is reduced due to the 
competition between $\eta$ and $\omega$, which is similar to Fig.\ref{fg:cden-zeroa}.  For higher chemical potential, 
nonzero $\omega$ with vanishing $\delta$ corresponds to the solution with $h_L=h_R$ at $a/f_a=\pi$, which is in 
agreement with the result obtained in \cite{Murgana:2024djt}. The temperature dependence of $\omega$ for fixed 
$\mu=350$ and $420\,\text{MeV}$ at $a/f_a=\pi$ is displayed in Fig.\ref{wtpi}.  
      
\begin{figure}[ht]
   \begin{center}
   \includegraphics[scale=0.35]{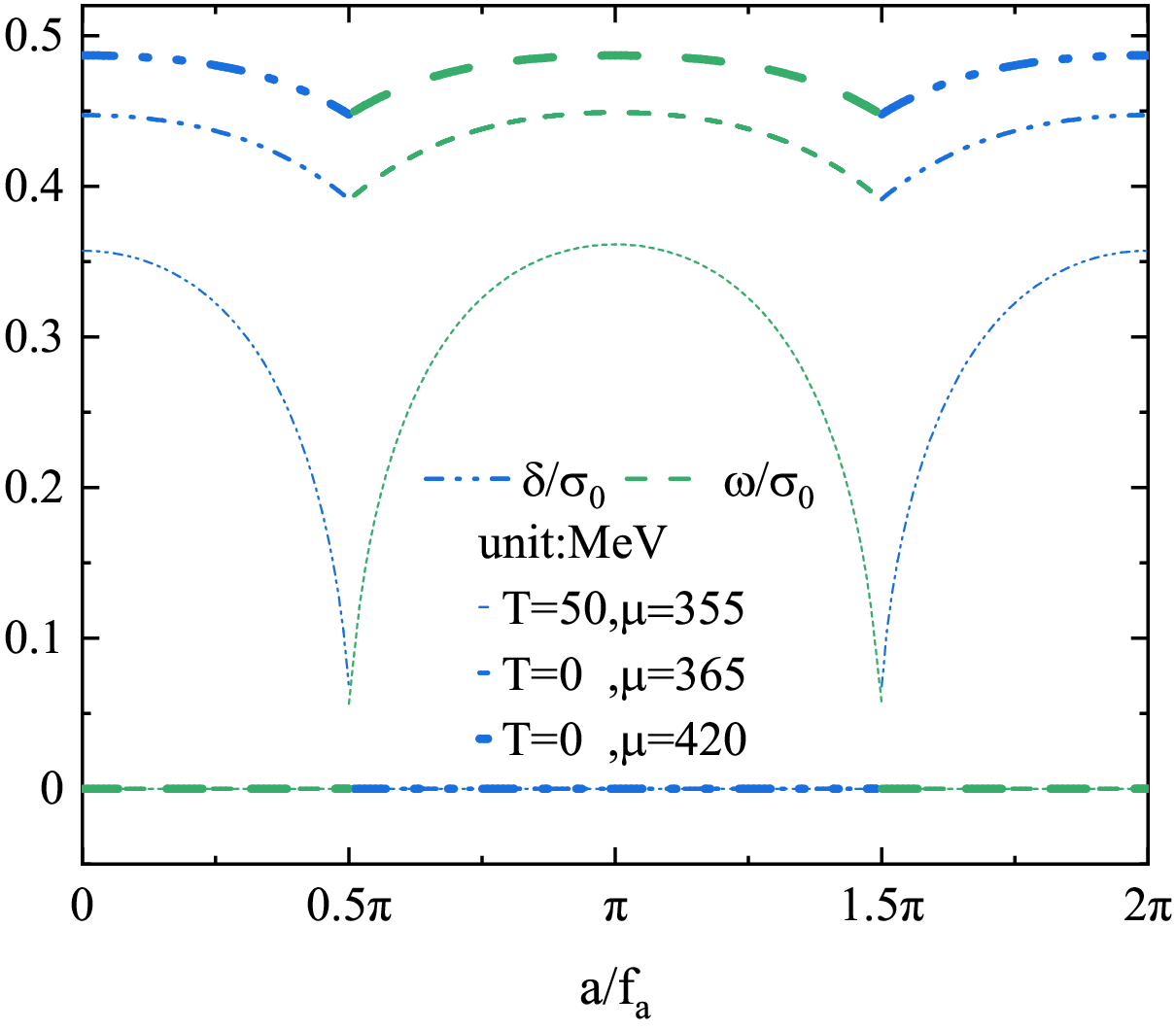}
   \includegraphics[scale=0.35]{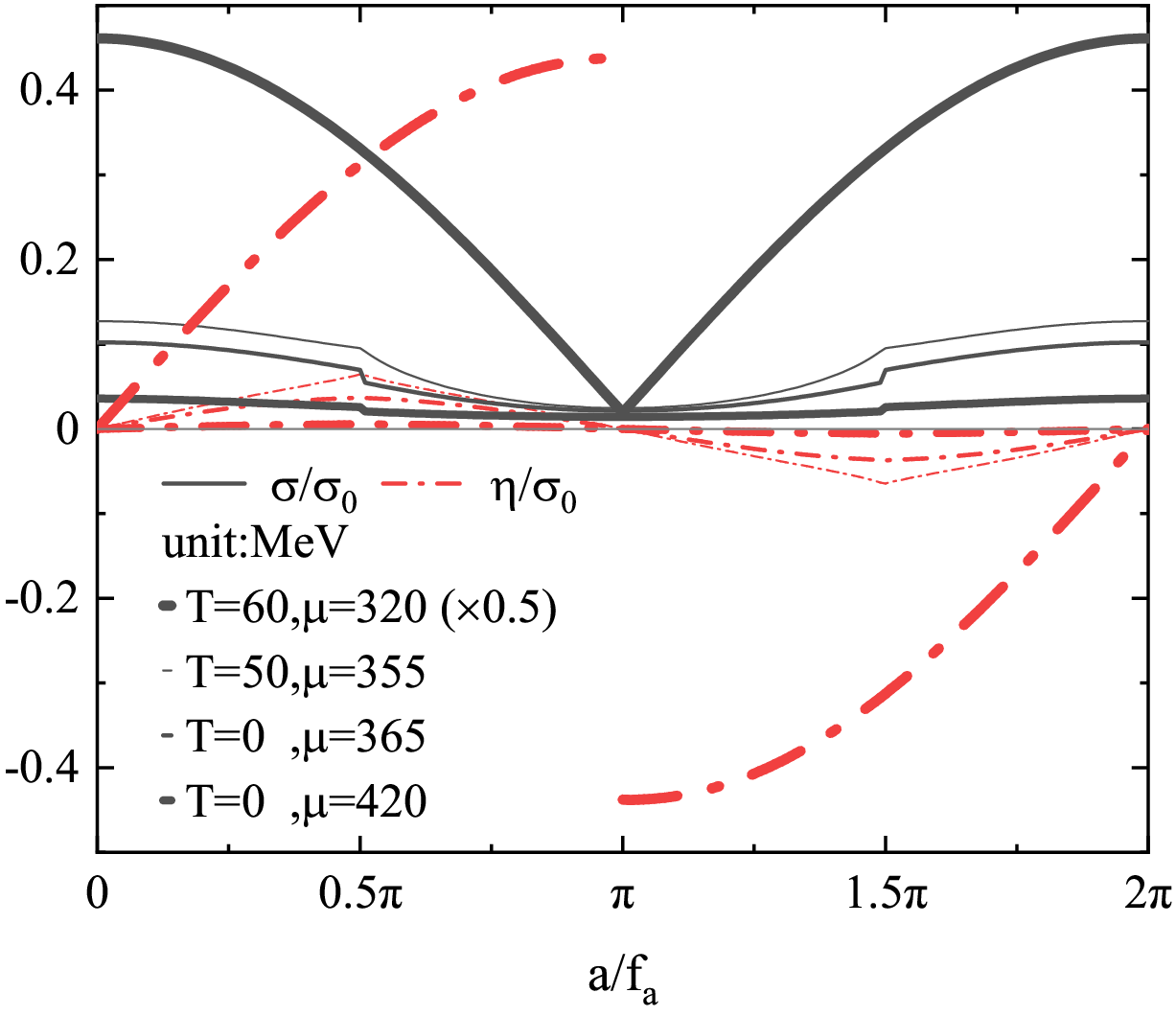}
   \end{center}
   \caption{ The normalized diquark condensates $\delta$ and $\omega$ (upper) and quark-antiquark condensates 
   $\sigma$ and $\eta$ (lower) versus $\theta=a/f_a$ for several $(T,\mu)$ points. The parameter $c$ is fixed 
   as 0.2.}
 \label{fg:condenVsa}
\end{figure}

In Fig.\ref{fg:condenVsa}, we demonstrate $\sigma$, $\eta$, $\delta$, and $\omega$ as functions 
of $\theta$ for several $(T,\mu)$ points. The upper panel shows that at $(T,\mu)=(0,420)$, 
$\delta$ appears but $\omega$ vanishes in the $\theta$ ranges $(0,\pi/2)$ and $(3\pi/2,2\pi)$; while 
in the $\theta$ range $(\pi/2,3\pi/2)$, $\omega$ emerges and $\delta$ vanishes. This implies a first order 
phase transition happens at $\theta=\pi/2$ ($3\pi/2$), where the scalar (pseudoscalar) diquark condensate 
changes into the pseudoscalar (scalar) one. This result is consistent with what obtained in \cite{Murgana:2024djt}. 
Similar phase transitions are observed at $(T,\mu)$=$(0,365)$ and $(50,355)$, where diquark condensates 
are weakened due to the decrease of $\mu$ and/or the increase of $T$. The quark-antiquark condensates 
$\sigma$ and $\eta$ at the same $(T,\mu)$ points are displayed in the lower panel. We see that for the 
point $(T,\mu)=(0,420)$, which is a bit far away from the phase boundary at zero $\theta$, the magnitudes 
of $\sigma$ and $\eta$ are obviously less than that of the nonzero $\delta$ or $\omega$. For points 
$(T,\mu)$=$(0,365)$ and $(50,355)$, which are close to the low temperature phase boundary at zero $\theta$, 
the difference between the quark-antiquark condensates and the non-vanishing diquark condensate $\delta$ 
or $\omega$ becomes smaller. In all the three cases, the maximum of the magnitude of $\omega$ locates at 
$\theta=\pi$, where the magnitude of $\sigma$ becomes smallest and $\eta$ is zero. Note that the Dashen's 
phenomenon only appears at lower $T$ and $\mu$ where the CS does't emerge, as indicated by the point 
$(T,\mu)=(60,320)$ in the lower panel. 

\begin{figure}[ht]
  \begin{center}
  \includegraphics[scale=0.35]{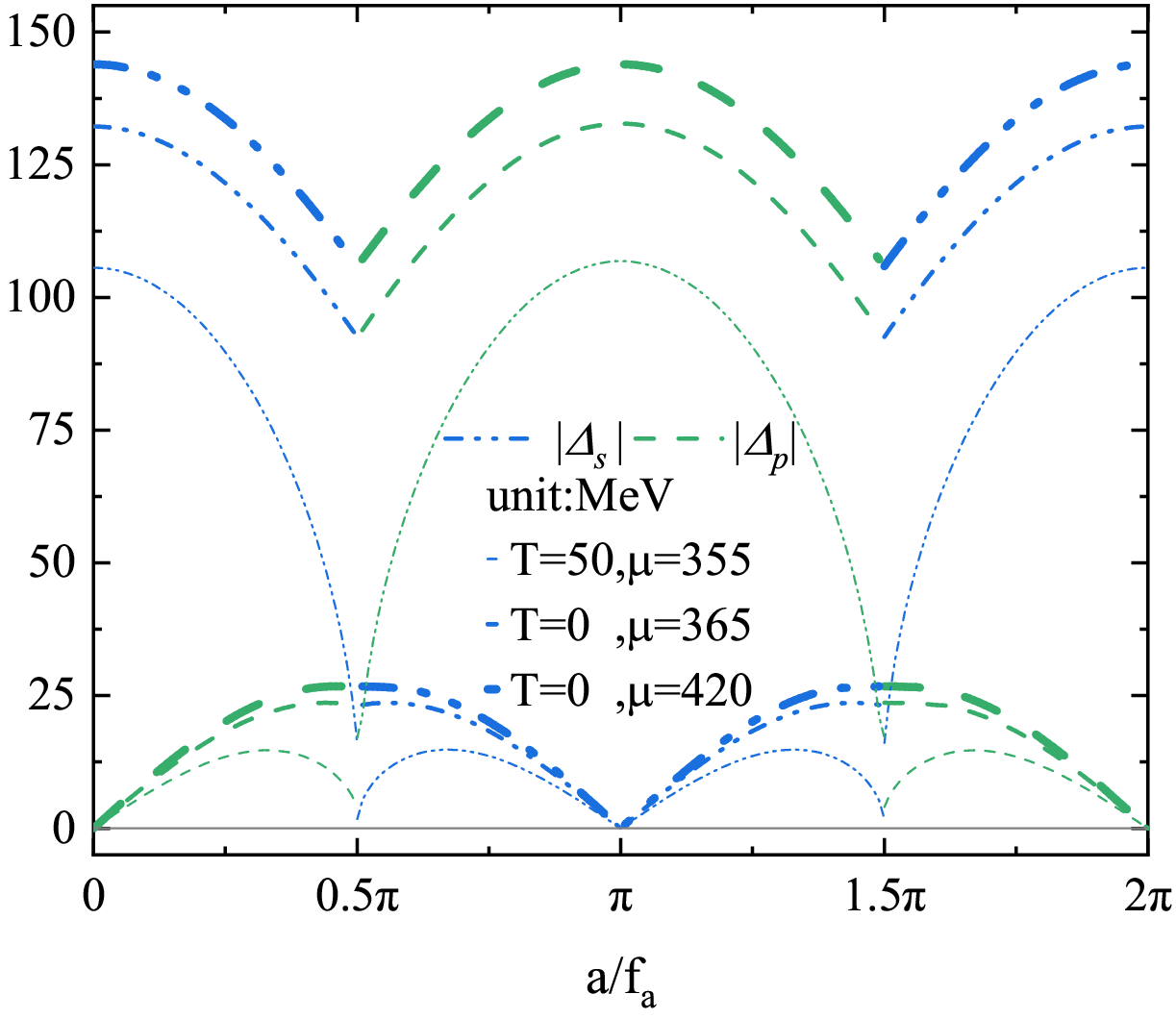} 
  \includegraphics[scale=0.35]{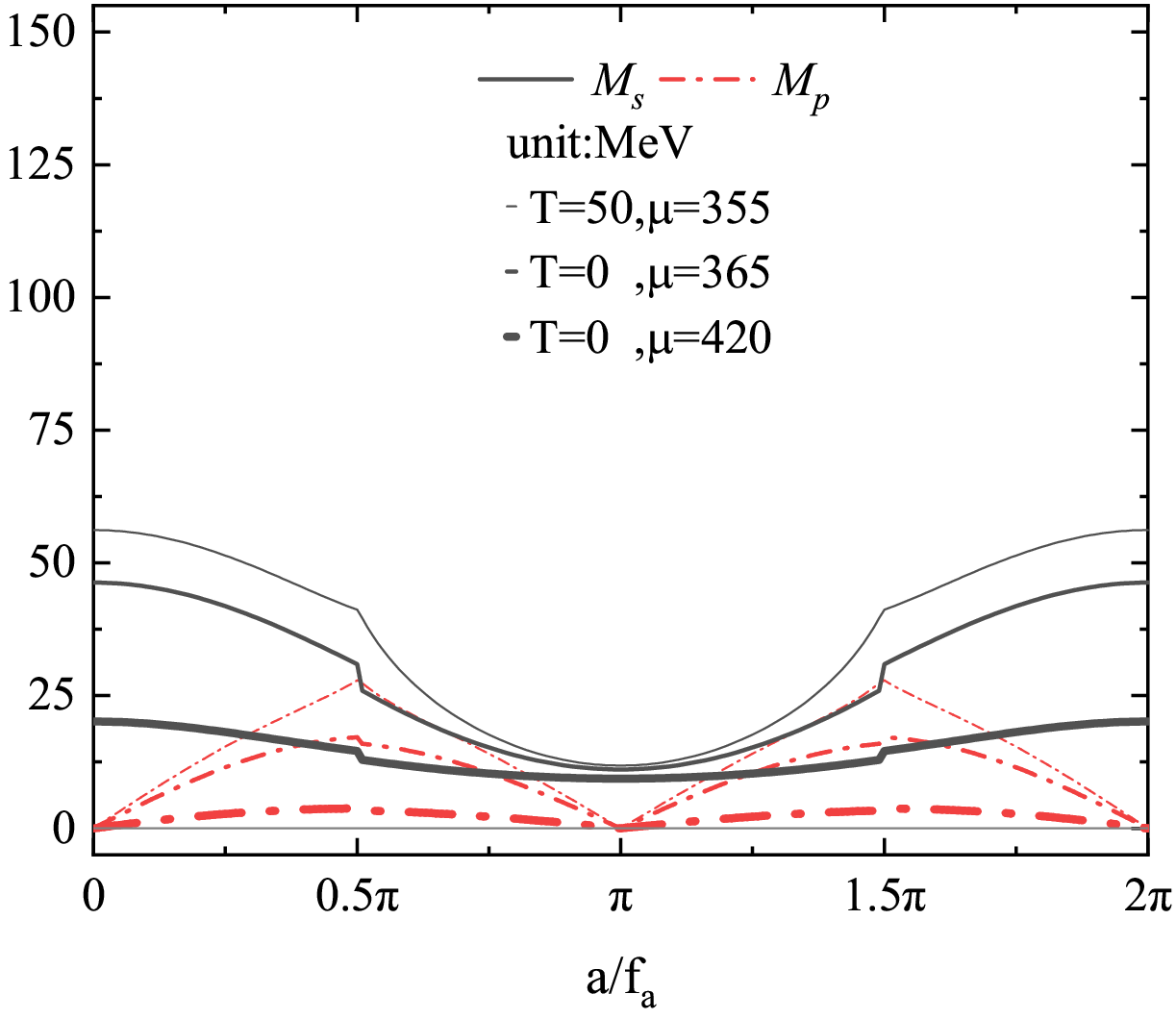}
  \end{center} 
   \caption{ Majorana masses $|\Delta_s|$ and $|\Delta_p|$ (upper) and Dirac masses $M_s$ and $M_p$ (lower) 
   versus $\theta=a/f_a$. The upper (lower) panel is plotted under the same conditions as that in the 
   upper (lower) panel of Fig.\ref{fg:condenVsa}. The parameter $c$ is fixed as 0.2.}
   \label{fg:gapsVsa}
\end{figure}

Figure \ref{fg:gapsVsa} shows the energy gaps $M_s$, $M_p$, $|\Delta_s|$, and $|\Delta_p|$ versus $\theta$=$a/f_a$ 
under the same conditions as that in Fig.\ref{fg:condenVsa}. Since $\Delta_s$ and $\Delta_p$ contain 
contributions from both $\delta$ and $\omega$, both the gaps form in the whole range $\theta$=$(0,2\pi)$.
Because Fig.\ref{fg:gapsVsa} is symmetrical about the vertical axis, we only concentrate on the left part 
of each panel. The upper panel shows $|\Delta_s|$ ($|\Delta_p|$) decreases (increases) with $\theta$ in the 
range $\theta$=$(0,\pi/2)$ and $|\Delta_p|$ ($|\Delta_s|$) increases (decreases) in $\theta$=$(\pi/2,\pi)$. 
According to Eq.\eqref{eq:PMjMass},  the pseudoscalar Majarona mass becomes $|\Delta_p|=|2H_2 \delta \sin{(a/f_a)}|$ 
in $\theta$=$(0,\pi/2)$ since $\omega$ is zero. This formula can be used to explain the behavior of $|\Delta_p|$ 
in this range: For $(T,\mu)$=$(0,420)$, the condensate $\delta$ decreases relatively slowly in $\theta$=$(0,\pi/2)$ 
(see Fig.\ref{fg:condenVsa}) and thus $|\Delta_p|$ increases with $\theta$ in this range as $\sin{(a/f_a)}$ 
does; For $(T,\mu)$=$(50,355)$, $|\Delta_p|$ first increases and then decreases in $\theta$=$(0,\pi/2)$ 
because $\delta$ drops rapidly near the left side of $\theta=\pi/2$. The behavior of $|\Delta_s|$ in 
$\theta$=$(\pi/2,\pi)$ can be understood in a similar way according to Eq.\eqref{eq:SMjMass}. The lower 
panel indicates that $M_s$ monotonically decreases with $\theta$ up to $\pi$, while $M_p$ first 
increases with $\theta$ up to $\pi/2$ and then decreases up to $\pi$ at which it becomes zero. We see 
that $|\Delta_s|$ ($|\Delta_p|$) is much lager than the other three gaps in $\theta=[0,\pi/2]$ ($\theta=[\pi/2,\pi]$) 
for the two zero temperature points $(T,\mu)$=$(0,420)$ and $(0,365)$. But for the point $(T,\mu)$=$(50,355)$ 
at a relatively higher temperature, though $|\Delta_s|$ ($|\Delta_p|$) near $\theta=0$ $(\pi)$ is still 
obviously larger than $M_s$, it becomes comparable to $M_s$ and $M_p$ around $\theta=\pi/2$ and $3\pi/2$.         

\subsection{Axion potential at finite $T$ and $\mu$ \label{subs:potential}}

\begin{figure}[ht!]
   \begin{center}
   \includegraphics[scale=0.35]{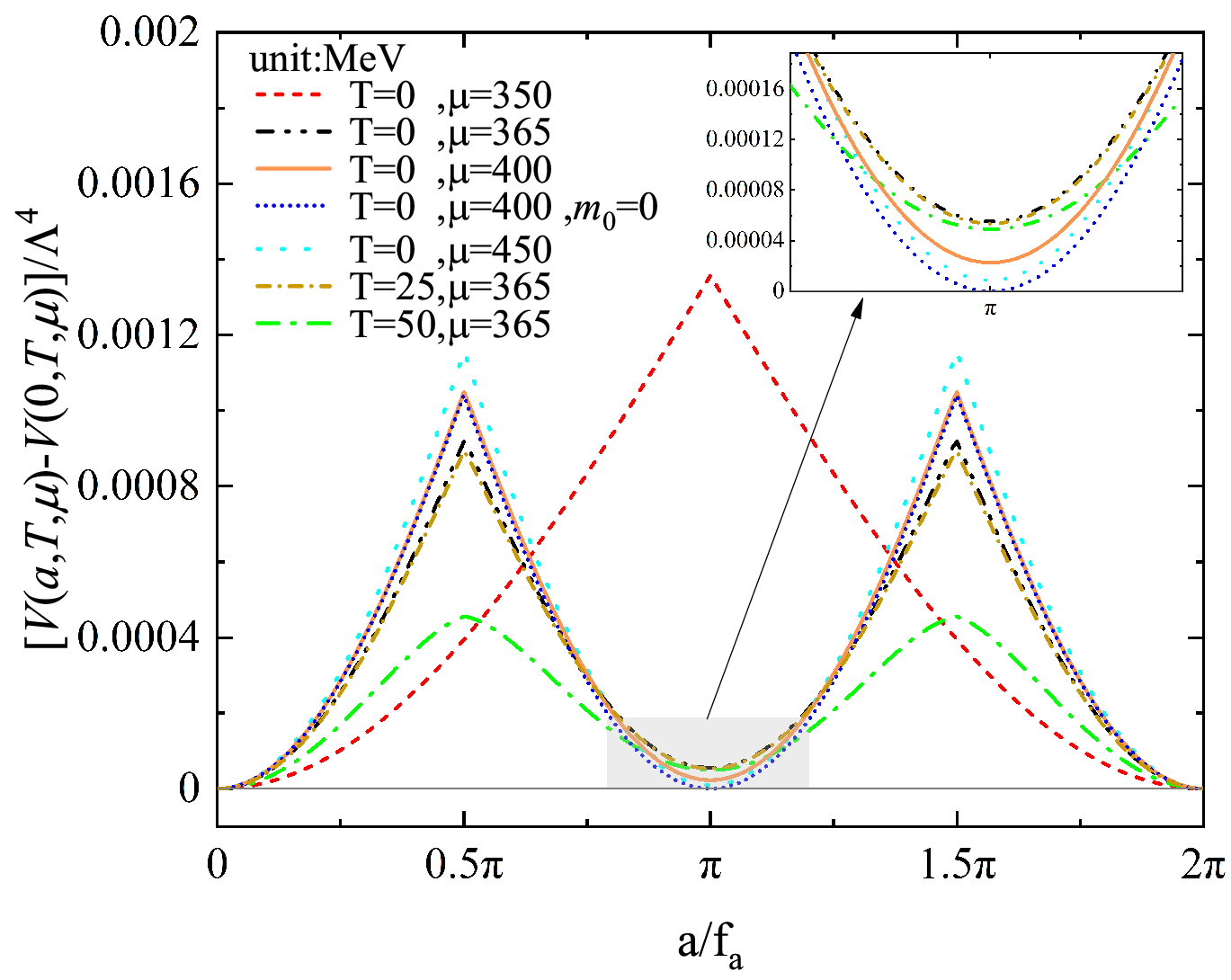}
   \end{center}
   \caption{Axion potential $V(a/f_a,T,\mu)-V(0,T,\mu)$ as functions of $\theta=a/f_a$ for several $(T,\mu)$ 
   points. The potential is measured in units of ${\Lambda ^4}$ with $\Lambda  = 590{\rm{MeV}}$. The parameter $c$
   is fixed as 0.2.}
   \label{fg:zhouqi}
\end{figure}

The axion potential $V(a,T,\mu)-V(0,T,\mu)$ versus $\theta \equiv a/f_a$ at different ($T$,$\mu$) points is 
shown in Fig.\ref{fg:zhouqi}. We see that the behaviors of the axion potential with and without the CS are quite 
different. For ($T$,$\mu$)=$(0,350)$, the diquark condensate does't form and the axion potential exhibits only 
peak at $\theta=\pi$ in the range $\theta=[0,2\pi]$. However, for other cases with the 2CS, the axion potential 
displays two degenerate peaks at $\theta=\pi/2$ and $3\pi/2$, respectively. Moreover, the axion potential at 
$\theta=\pi$ becomes a local minimum rather than a maximum in the presence of CS. By comparing the cases of 
($T,\mu$)=$(0,365)$, $(0,400)$, and $(0,450)$, we see that the axion potential becomes larger with $\mu$; 
by comparing the points of ($T,\mu$)=$(0,365)$, $(25,365)$, and $(50,365)$, which are close to the phase 
boundary at zero $\theta$, we observe that the axion potential gets smaller with $T$.    

Figure \ref{fg:zhouqi} shows that the axion potential exhibits an approximate periodicity with the period $\pi$ 
due to the presence of 2CS. Such an approximate period turns into an exact one in the chiral limit, as indicated by 
the blue doted-line for ($T$,$\mu$)=$(0,400)$. The emergence of the period $\pi$ is in agreement with the result 
obtained in \cite{Murgana:2024djt} where Dirac type masses are not included. Actually, the reason for the period 
$\pi$ is the same as that given in \cite{Murgana:2024djt} since both $\sigma$ and $\eta$ are dynamically absent 
for $m_0=0$ in our calculations. In other words, the period $\pi$ obtained in \cite{Murgana:2024djt} is broken 
by the current quark masses. We see that such a breaking is quite small at lower temperature and larger quark chemical 
potential since the axion potentials calculated with and without $m_0$ are almost coincident. Of course, the breaking 
becomes relatively obvious near the phase boundary in the $T$-$\mu$ phase diagram because of the enhanced $\sigma$ 
and $\eta$, which can be judged by the difference between the axion potentials at $\theta=\pi$ and zero (see the subgraph). 
We can expect that the breaking of the periodicity with the period $\pi$ will become more seriously if the coupling 
constants in diquark channels are weak enough.      

\subsection{Topological susceptibility, axion mass, and self-coupling \label{subs:axion}} 

In this subsection, we present our numerical results on the topological susceptibility, axion mass, and 
axion self-coupling in the presence of the condensates $\sigma$, $\eta$, $\delta$, and $\omega$.  

The topological susceptibility is a parameter that characterizes the response of QCD vacuum to topological 
charge fluctuations, which is defined as 
\begin{equation}\label{eq:tosus}
  \chi_{t}=\frac{d^2\Omega}{d\theta^2}\Big|_{\theta=0}. 
\end{equation} 
According to Eq.\eqref{eq:amass}, this quantity is proportional to the axion mass squared. The evaluation 
of $\chi_t$ at vacuum and medium has been performed using different non-perturbative methods such as LQCD, 
$\chi PT$ and effective models \cite{Borsanyi:2016ksw,Bonati:2015vqz,Petreczky:2016vrs,GrillidiCortona:2015jxo,Lu:2018ukl}.

The calculation of $\chi_t$ in the 2CS phase has been given in \cite{Murgana:2024djt}, where an analytical 
formula has been derived which takes the form  
\begin{equation}\label{eq:tsformula}
  \chi_{t}=H_2 \delta^2 \frac{1+\frac{H_2}{H_1}}{1-\frac{H_2}{H_1}}=H_2 \delta^2(1-2c). 
\end{equation}
The rightmost side of \eqref{eq:tsformula} is obtained using the relation $H_2/H_1=G_2/G_1=c/(1-c)$, which holds 
under the assumption that the coupling ratios $H_1/G_1$ and $H_2/G_2$ are both limited to the Fierz transformations.
Note that the analytical formula \eqref{eq:tsformula} does't hold when the quark masses are considered. 
Here we will focus on how $\chi_t$ is affected by the chiral condensate in the presence of the 2CS, 
especially how it varies cross the chiral phase transition line. 

\begin{figure}[ht]
   \begin{center}
   \includegraphics[scale=0.35]{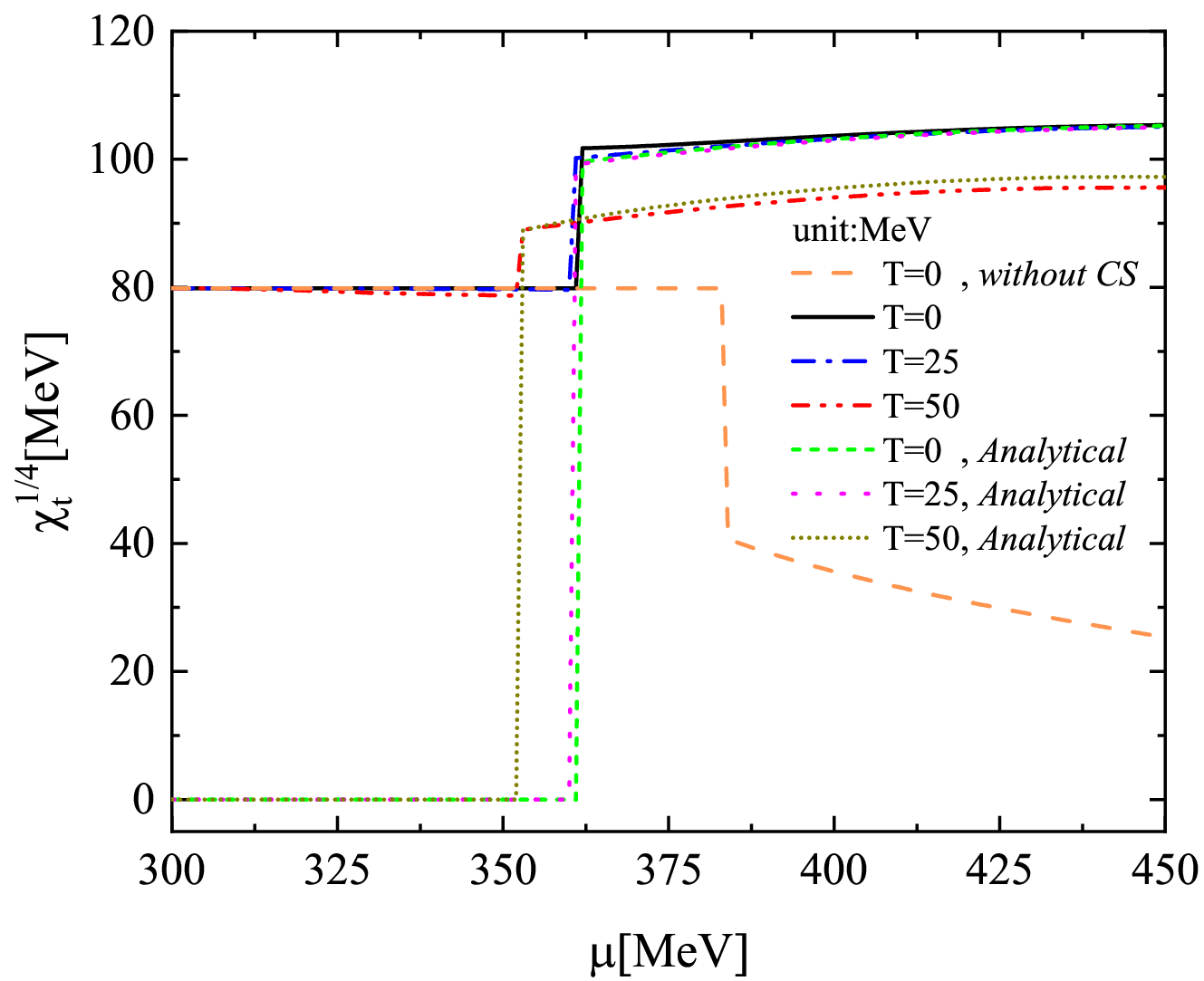}
   \end{center}
   \caption{Topological susceptibility $\chi _t^{1/4}$ versus $\mu$ for several values of $T$. The parameter $c$
   is fixed as 0.2.}
   \label{fg:topVst}
\end{figure}

In Fig.\ref{fg:topVst}, we show the topological susceptibility as functions of $\mu$ for different temperatures 
with $c=0.2$. For comparison, we also present the results without considering the 2CS and with the 2CS but using 
the analytical formula \eqref{eq:tsformula}. We see that before the phase transition, $\chi_t$ calculated at zero 
and low temperatures keeps almost unchanged with $\chi _t^{1/4}\approx 79.8\text{MeV}$, which is close to the 
standard vacuum value $\sim{77.8}\,\text{MeV}$ obtained in the $\chi\text{PT}$ method \cite{GrillidiCortona:2015jxo}.
At the critical chemical potential, $\chi_t$ increases suddenly ($\chi_t^{1/4}$ reaches $\sim100\text{MeV}$ at $T=0$) 
and then grows slowly with $\mu$. This is quite different from the case without including the CS where $\chi_t$ drops 
abruptly at the critical point \cite{Lu:2018ukl,Zhang:2023lij}, as indicated by the long dashed line for $T=0$ in Fig.\ref{fg:topVst}. 
On the other hand, the numerical results of $\chi_t$ in the 2CS phase at $T=0$ and $25\,\text{MeV}$ agree quite well with 
that calculated using the analytical formula for higher $\mu$ and small deviations appear near the phase transition point. 
This suggests that for higher $\mu$ and lower $T$ the dominant contribution to $\chi_t$ comes from the diquark condensate 
since the chiral condensate is suppressed significantly. Fig.\ref{fg:topVst} shows that the obvious deviation emerges at 
higher $\mu$ for $T=50\,\text{MeV}$. This can be attributed to the increased weight of the contribution from the chiral 
condensate due to the suppression of the diquark condensate at higher $T$.           
     
\begin{figure}[ht]
   \centering
   \includegraphics[scale=0.35]{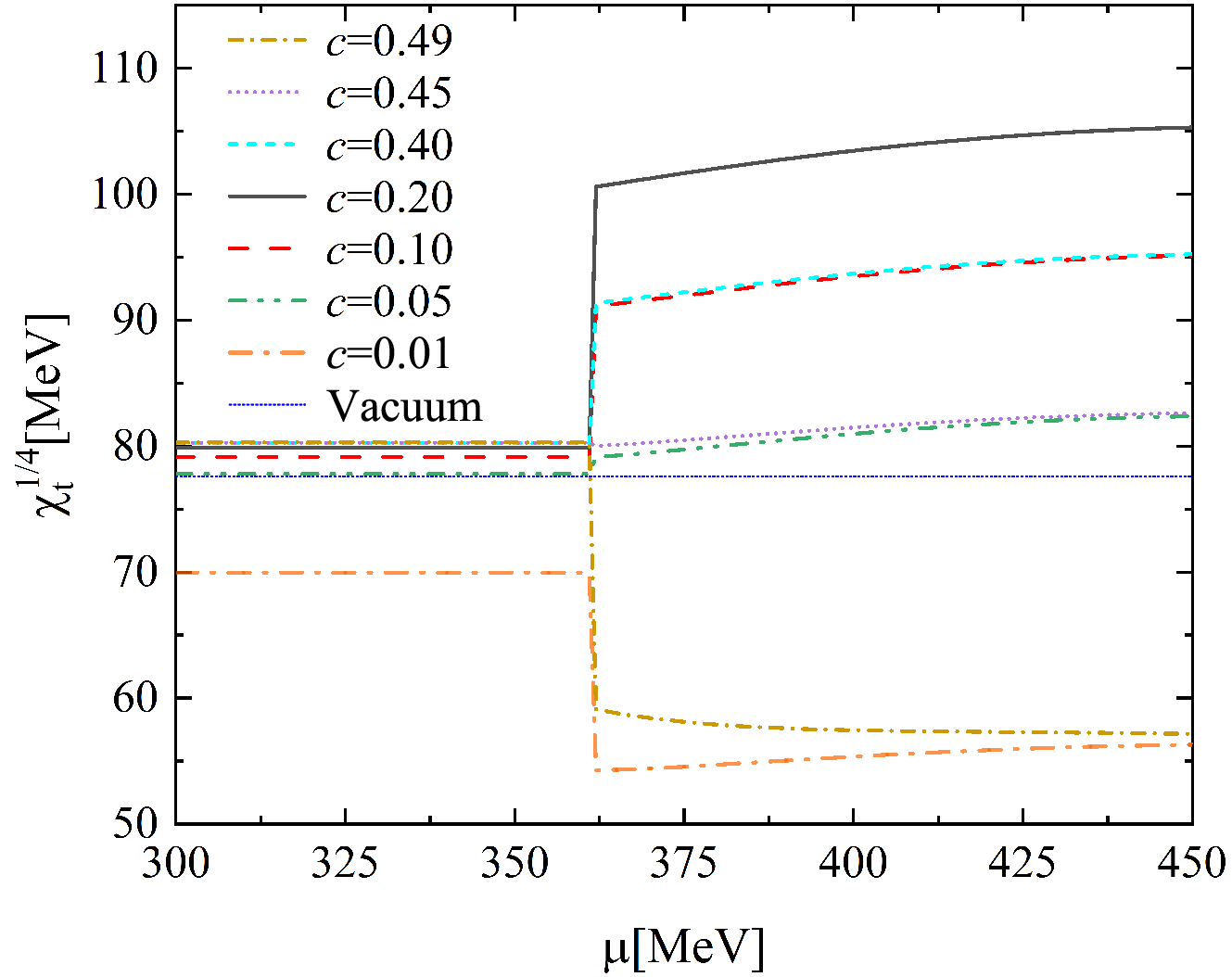}
   \includegraphics[scale=0.35]{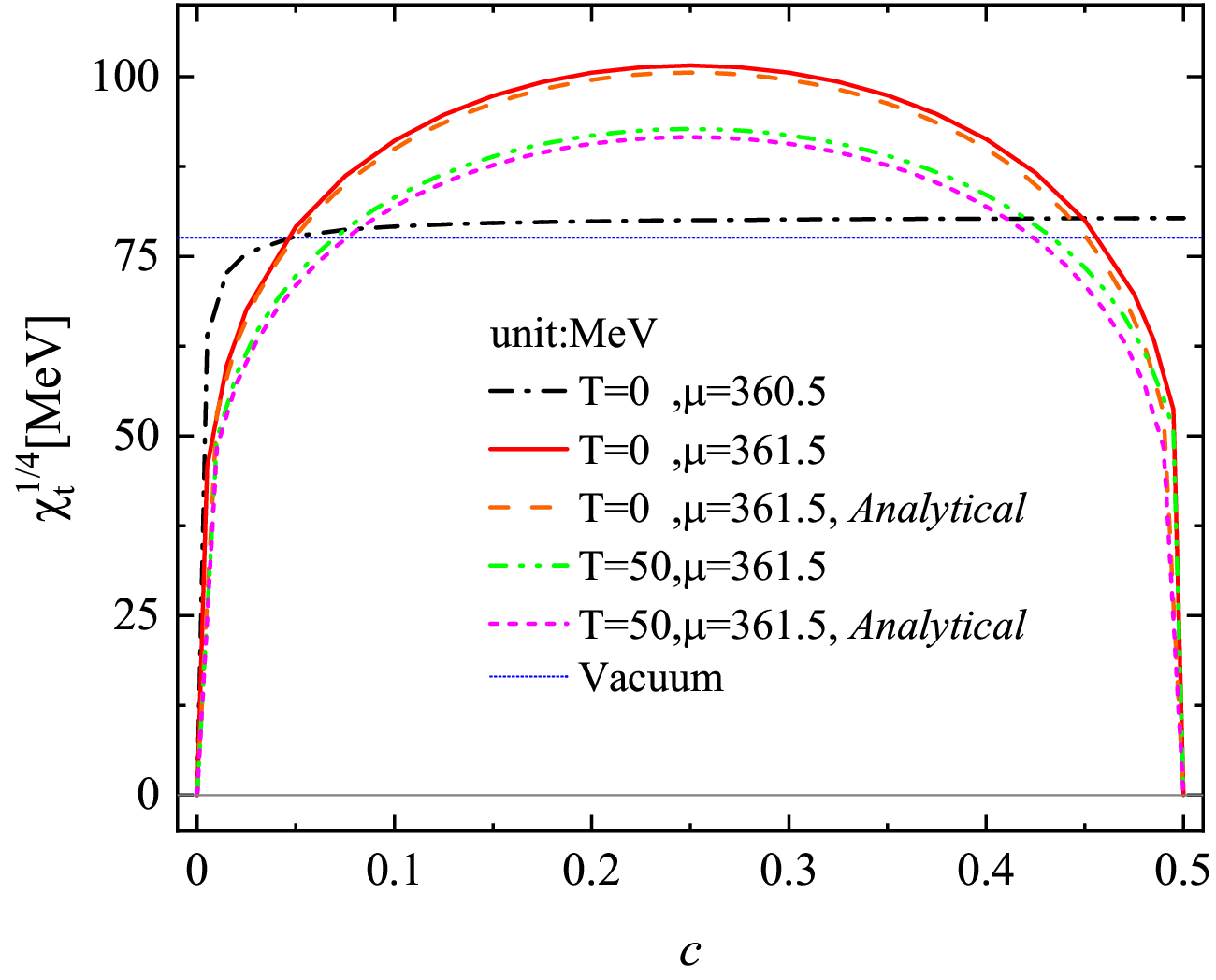}
   \caption{Upper panel: $\chi_t^{1/4}$ versus $\mu $ for several values of $c$  at $T=0$.
    Lower panel: $\chi _t^{1/4}$ versus $c$ for several $T$-$\mu$ points in the proximate of 
    the low temperature phase boundary. }
   \label{fg:topVsc}
\end{figure}

Figure \ref{fg:topVsc} displays how $\chi_t$ depends on the parameter $c$. According to Eq.\eqref{eq:tsformula}, 
$\chi_t$ will become zero for $c=0.5$. The numerical calculation indicates that $\chi_t$ or $m_a^2$ becomes 
negative when $c>0.5$ if $\mu$ is large enough. So we only consider the range $c=(0,0.5)$
(the upper bound 0.5 will be sightly modified for the case with massive quarks.). The upper panel
shows that the first order phase transitions with different $c$ almost happen at the same critical chemical 
potential $\mu_c=361\,\text{MeV} $ for $T=0$. The abrupt increase of $\chi_t$ due to the emergence of the 
2CS appears roughly in the range $(0.05,0.45)$: for $\mu<\mu_c$ , $\chi_t^{1/4}$ is insensitive to $c$, 
but it becomes quite sensitive to $c$ for $\mu>\mu_c$. For a very small $c=0.01$, $\chi_t^{1/4}$ is 
obviously less than the standard vacuum value for $\mu<\mu_c$ and it further drops in the 2CS phase. 
For $c=0.49$, even $\chi_t^{1/4}$ is still close to the standard vacuum value for $\mu<\mu_c$, it also 
decreases significantly at the chiral transition point due to the presence of CS. In the lower panel, we show 
$\chi_t$ versus $c$ in the range $c=(0,0.5)$ for three ($T,\mu$) points near the phase boundary. In the 
chiral symmetry breaking phase with ($T,\mu$)= $(0,360.5)$, the CS does't appear and $\chi_t^{1/4}$ keeps 
almost unchanged for $c>0.05$; but for $c<0.05$, it drops obviously with the decreasing of $c$. In the CS phase 
with ($T,\mu$)= $(0,361.5)$, $\chi_t^{1/4}$ increases with $c$ up to $\sim0.25$ and then decreases up to 
$\sim0.5$. Comparing to its vacuum value,  we see that $\chi_t^{1/4}$ become larger in the range $c=(0.05,0.45)$ 
for $T=0$ due to the presence of 2CS. The similar conclusion is obtained for $T=50\,\text{MeV}$ where the range 
for the enhancement of $\chi_t$ due to the CS is shortened. For comparison, we also report $\chi_t^{1/4}$ obtained 
using the analytic formula \eqref{eq:tsformula} in the 2CS phase for $T=0,\,50\,\text{MeV}$. We see that the 
deviation is quite small and thus the analytic formula is still a good approximation for the calculation of 
$\chi_t$ in the presence of $\sigma$ and $\delta$.  
 
\begin{figure}[ht]
   \begin{center}
   \includegraphics[scale=0.35]{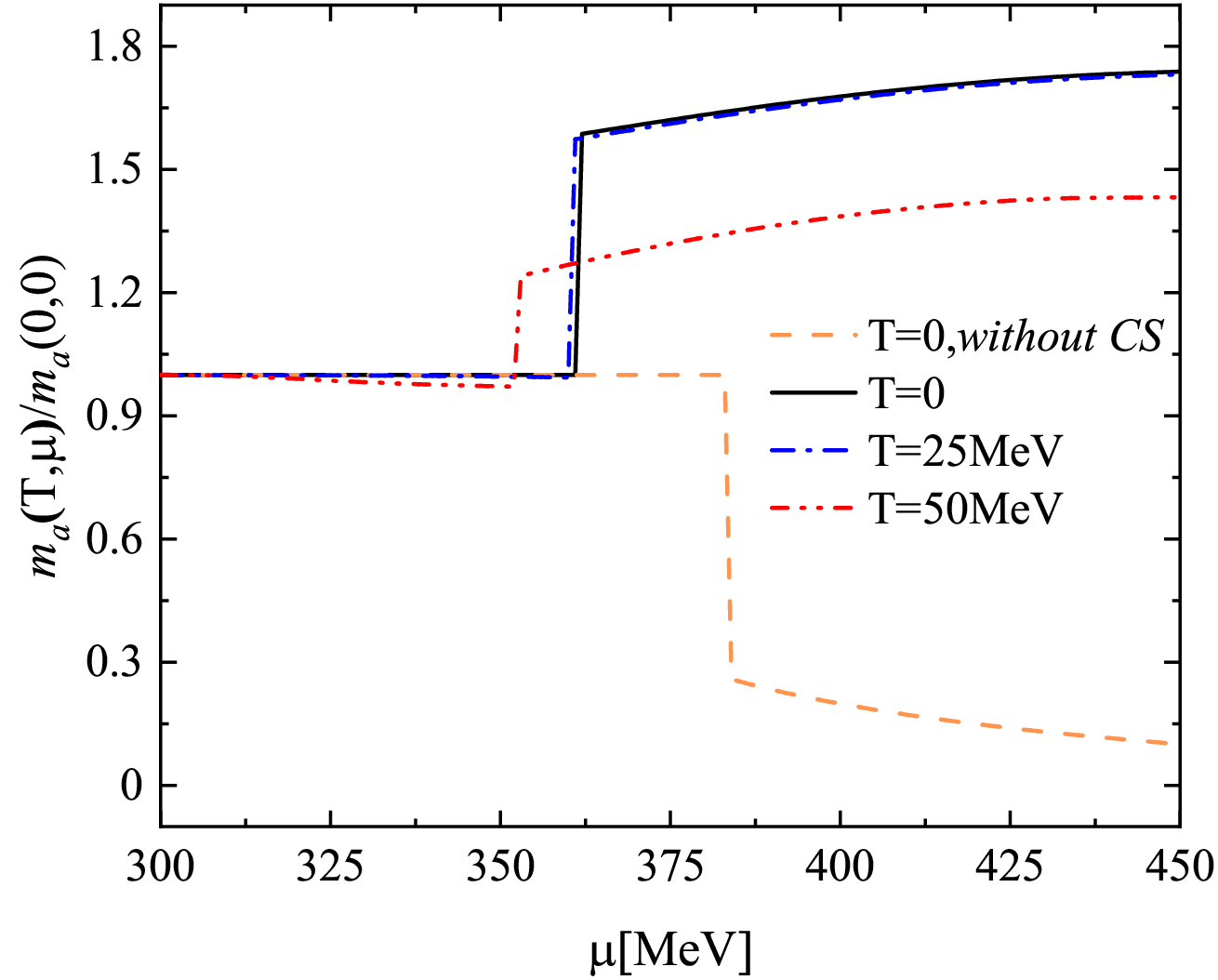}
   \end{center}
   \caption{The normalized axion mass ${m_a}$  with $c=0.2$ as a function of $\mu$ under the same conditions 
   as that in Fig.\ref{fg:topVst}. }
   \label{fg:massVst}
 \end{figure}

\begin{figure}[ht]
   \centering
   \includegraphics[scale=0.35]{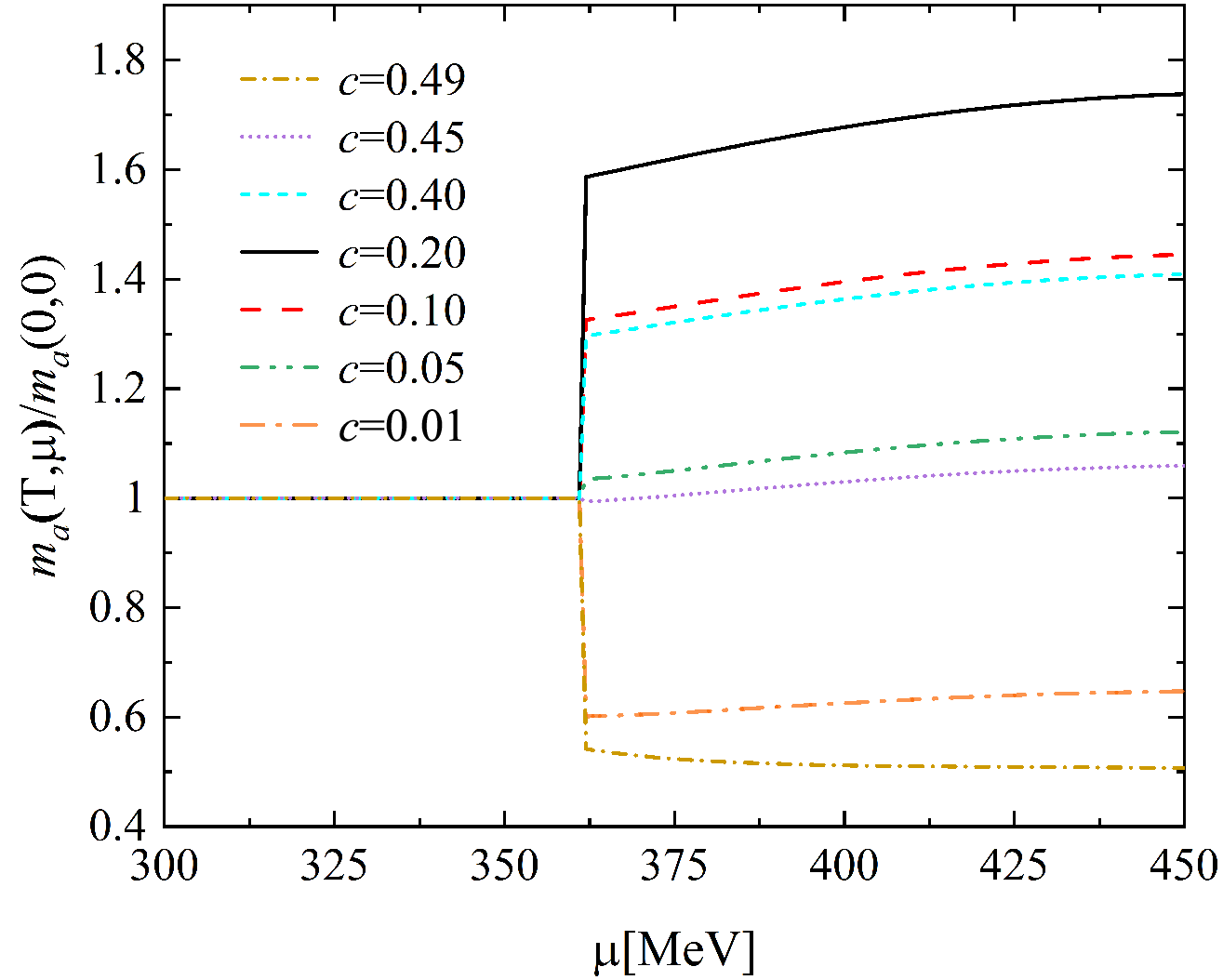}
   \caption{The axion mass ${m_a}$ (normalized by its vacuum value obtained using the same $c$) as a function 
   of $\mu$ under the same conditions as that in the upper panel of Fig.\ref{fg:topVsc}.}
   \label{fg:massVsc}
\end{figure}

The normalized axion mass as the function of $\mu$ under the same conditions as that in Fig.\ref{fg:topVst} 
and the upper panel of Fig.\ref{fg:topVsc} are shown in Figs.\ref{fg:massVst} and \ref{fg:massVsc}, respectively. 
Since the axion mass squared is proportional to $\chi_t$, $m_a$ also rises suddenly at the chiral phase transition 
point due to the appearance of the CS and then increases with $\mu$ if $c$ is not very close to zero or 0.5. 
This is distinct with the case without the CS where $m_a$ decreases obviously at the chiral critical point, as 
indicated by the long dashed line in Fig.\ref {fg:massVst}. Similarly, Fig.\ref{fg:massVsc} displays that for  
$\mu<\mu_c$ ( $\mu>\mu_c$), the normalized $m_a$ is insensitive (sensitive) to $c$.
    
\begin{figure}[ht]
   \begin{center}
   \includegraphics[scale=0.35]{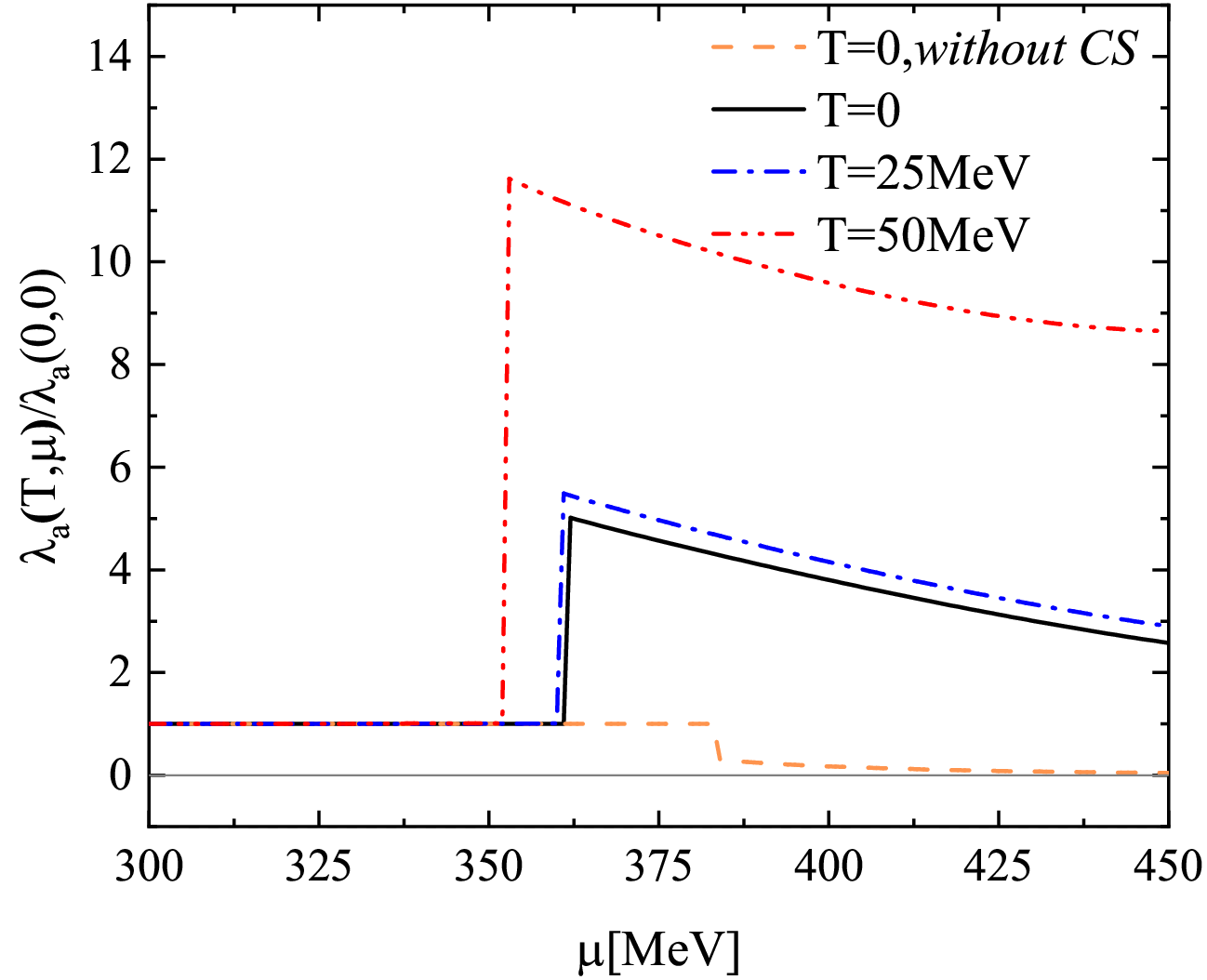}
   \end{center}
   \caption{The normalized axion self-coupling as a function of $\mu$ for several values of $T$ with $c=0.2$. }
   \label{fg:selfcVst}
\end{figure}

\begin{figure}[ht]
   \centering
    \includegraphics[scale=0.35]{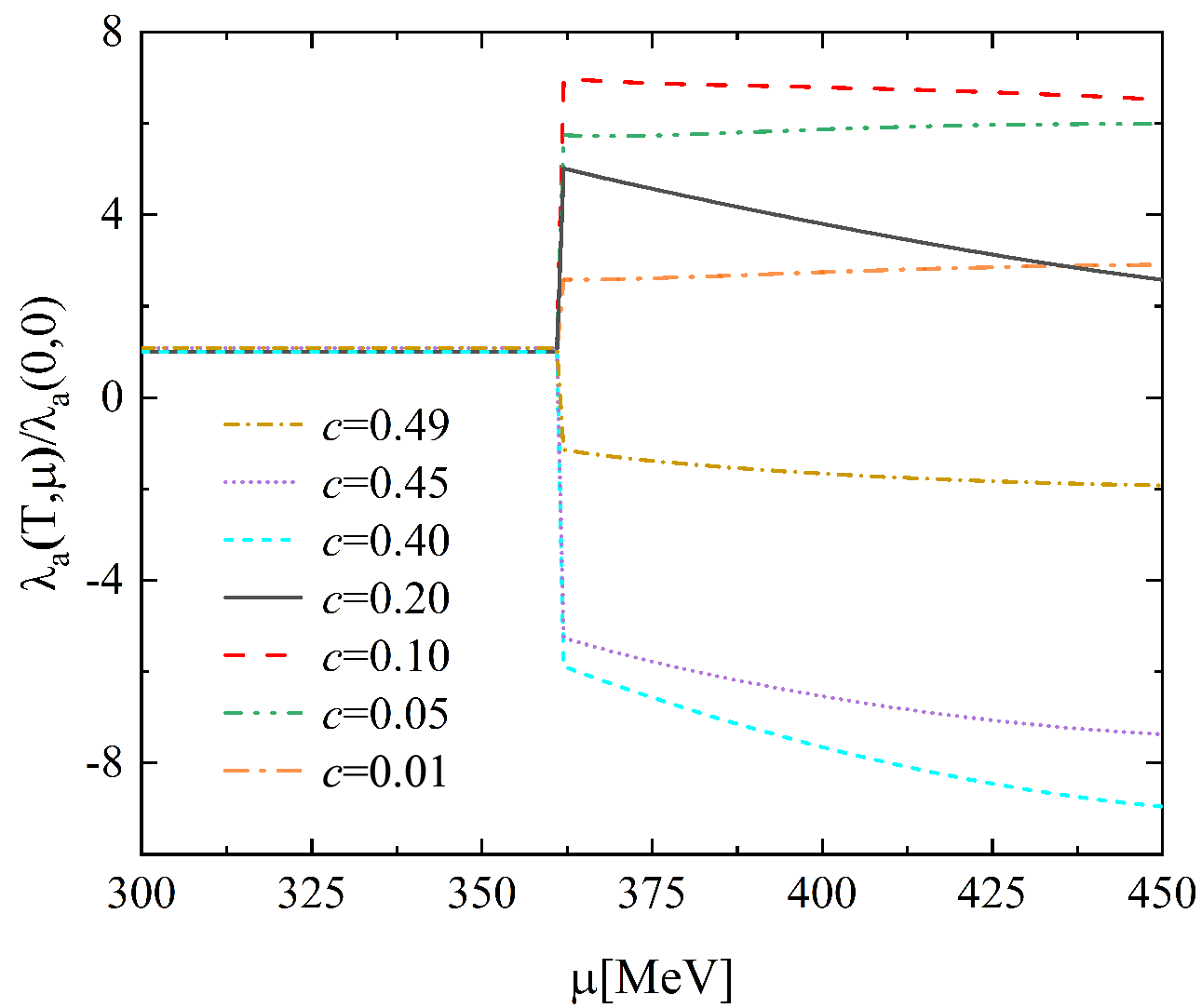}
    \includegraphics[scale=0.35]{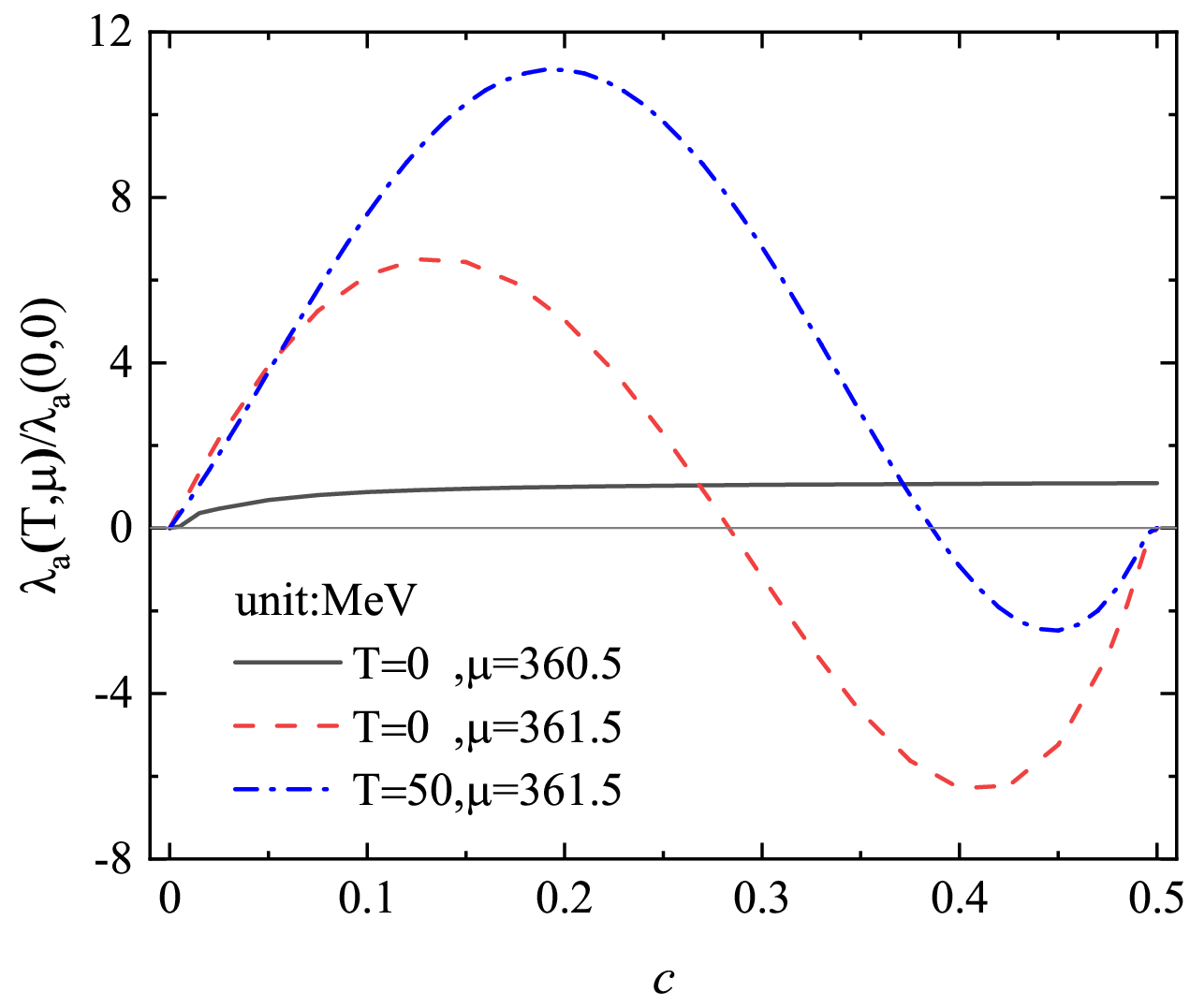}
   \caption{The normalized axion quartic self-coupling as functions of $\mu$ for 
   several values of $c$ at $T=0$ (upper) and $c$ at three $(T,\mu)$ points (lower). In the lower panel,  
  the chiral symmetry is broken at the point $(T,\mu)$=$(0,360.5)$ where the 2CS doesn't appear and restored 
  with the emergence of 2CS at $(T,\mu)$=$(0,361.5)$ and $(50,361.5)$ .}
   \label{fg:selfcVsc}
\end{figure}

The axion self-coupling $\lambda_a(T,\mu)$ (normalized by $\lambda_a(0,0)$) versus $\mu$ for $T=0,\,25,\,50\,\text{MeV}$ 
with $c=0.2$ is shown in Fig.\ref{fg:selfcVst}. At $T=0$, the normalized self-coupling rises significantly at $\mu_c$ 
and then decreases with $\mu$. This is also quite different from the case without the CS where the normalized self-coupling  
drops abruptly at the phase transition point, as indicated by the long dashed line. The similar behavior is observed at 
finite $T$ in the presence of the CS, where the magnitude of $\lambda_a(T,\mu)$ increases with $T$. In Fig.\ref{fg:selfcVsc}, 
we display the $c$ and $\mu$ dependences of $\lambda_a(T,\mu)/\lambda_a(0,0)$ under the same conditions as that in Fig.\ref{fg:topVsc}. 
We see that the $\mu$-dependence of the self-coupling is insensitive to $c$ in the chiral symmetry breaking phase but 
quite sensitive to it in the CS phase. Especially, the self-coupling becomes positive in the range $0.28<c<0.50$ ($0.38<c<0.50$) 
at $(T,\mu)$=$(0,361.5)$ ($(T,\mu)$=$(50,361.5)$). This means that the axion quartic self-interaction may become 
repulsive in the 2CS phase if $c$ is large enough. The similar conclusion is also obtained in \cite{Murgana:2024djt}.     
  
Moreover, the topological susceptibility and the axion self-coupling versus $\mu$ at $T=0$ for different ratios of $H_1/G_1$ 
and $H_2/G_2$ beyond the Fierz transformations are given in Appendix.       

\subsection{The axion domain walls \label{subs:wall}} 

As shown in Fig.\ref{fg:zhouqi}, the axion potential has two successive vacua at $\theta=0$ and $\theta=2\pi$, 
which permits axion domain wall solution \cite{Sikivie:1982qv} to interpolate between them (details on the domain wall 
derivation see textbooks \cite{Shifman:2022shi,Weinberg:2012pjx,Nagashima:2014tva}). The axion domain wall in the medium 
of dense quark matter without the CS has been explored within the NJL formalism \cite{Zhang:2023lij}. Here we report our 
study on properties of axion domain wall in the presence of 2CS.        

Following \cite{Zhang:2023lij}, the field equation for the axion $a(x)=f_a \theta(x)$ takes the form  
\begin{equation}
	\partial_\mu\partial^\mu \theta + 
\frac{1}{f_a^2}
	\frac{\partial V(\theta)}{\partial \theta}=0,
	\label{eq:thetaFE}
\end{equation} 
where $V(\theta)$ is the axion potential. Eq.\eqref{eq:thetaFE} has the solitary domain wall solution
\begin{equation}
\theta(x,t)=\theta(x-vt),\label{eq:DW1}
\end{equation}
where $v$ is the propagation speed of the soliton. The field equation can be rewritten as \cite{Zhang:2023lij}   
\begin{equation}
(1-v^2) \theta_{\xi\xi} = \frac{1}{f_a^2}\frac{\partial V(\theta)}{\partial \theta},
\label{eq:DW2}
\end{equation}  
where $\xi=x-vt$. Multiplying both sides of Eq.\eqref{eq:DW2} by $\theta_\xi$ and integrating with the boundary 
conditions of $\theta\rightarrow 0$ and $\theta_\xi\rightarrow 0$ for $\xi\rightarrow\pm\infty$,  one can obtain
the kink and antikink solutions  
\begin{equation}
\frac{d\theta}{\sqrt{V(\theta)}}
=\pm\sqrt{\frac{2}{f_a^2(1-v^2)}}d\xi.
\label{eq:kink}
\end{equation} 
As in \cite{Zhang:2023lij}, we only consider the soliton at rest and thus we have $x\equiv\xi$. In this case, 
integrating both sides of \eqref{eq:kink}, one obtains  
\begin{equation}
	\int_{\pi}^{\theta(x)}
	\frac{d\theta}{\sqrt{V(\theta)}}
	=	\pm x \sqrt{\frac{2}{f_a^2}},
	\label{eq:soliton}
\end{equation} 
where the upper limit of the left integration corresponds to the soliton profile $\theta(x)$ which 
center is required to satisfy $\theta(0)=\pi$. Using an analytical cosine potential (see the end of this subsection), 
it has been shown in \cite{Zhang:2023lij} that the thickness of the axion wall is directly related to the 
axion mass: the larger the axion mass, the thicker the wall.

\begin{figure}[ht]
   \begin{center}
   \includegraphics[scale=0.35]{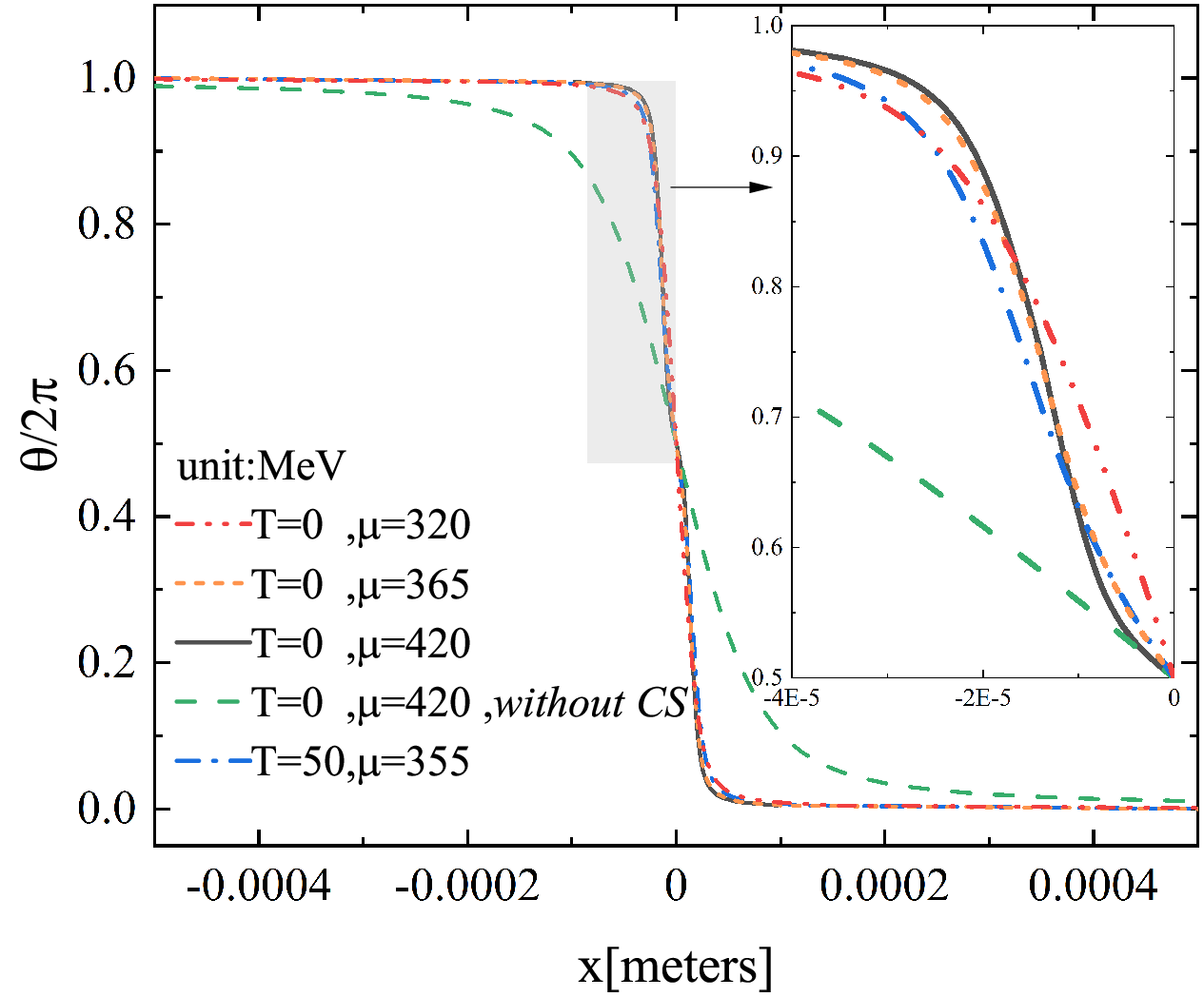}
   \end{center}
   \caption{Axion walls, $\theta = a/f_a$, versus $x$ at different ($T$,$\mu$) points for $c=0.2$. 
  ($T$,$\mu$)=(0,320) is a point in the chiral symmetry breaking phase. ($T$,$\mu$)=(0,365), (50,355) and (0,420) 
   are three points in the chiral restored phase with the 2CS: the former two (last one) are (is) close 
   to (far from) the phase boundary at $\theta=0$. }
   \label{fg:wallVSx}
\end{figure}
 
In Fig.\ref{fg:wallVSx}, we plot the axion walls as functions of $x$ at different $T$ and $\mu$. 
Following \cite{Zhang:2023lij}, $f_a=10^9 \text{GeV}$, which is located in the so called classical 
window, is adopted in the calculations. The red dash-doted-doted line corresponds to the wall in the  
chiral symmetry breaking phase at $(T,\mu)$=$(0,320)$ and the green dashed line the one in the chiral restored
phase without considering the CS at $(T,\mu)$=$(0,420)$. The left three lines are the axion walls in the 
presence of the 2CS which are obtained at $(T,\mu)$=$(0,420)$, $(0,365)$, and $(50,355)$ respectively. We 
see that the axion wall in the chiral restored phase without the 2CS is obviously wider than the other 
four cases. The reason can be traced back to the relatively large axion mass in the 2CS phase or in 
the chiral symmetry breaking phase. The inset indicates that the wall at $(T,\mu)$=$(0,420)$ is narrower 
than that at $(T,\mu)$=$(0,320)$ since the axion mass in the 2CS phase is larger than that in 
the chiral symmetry breaking phase for $c=0.2$.

\begin{figure}[ht]
   \begin{center}
   \includegraphics[scale=0.35]{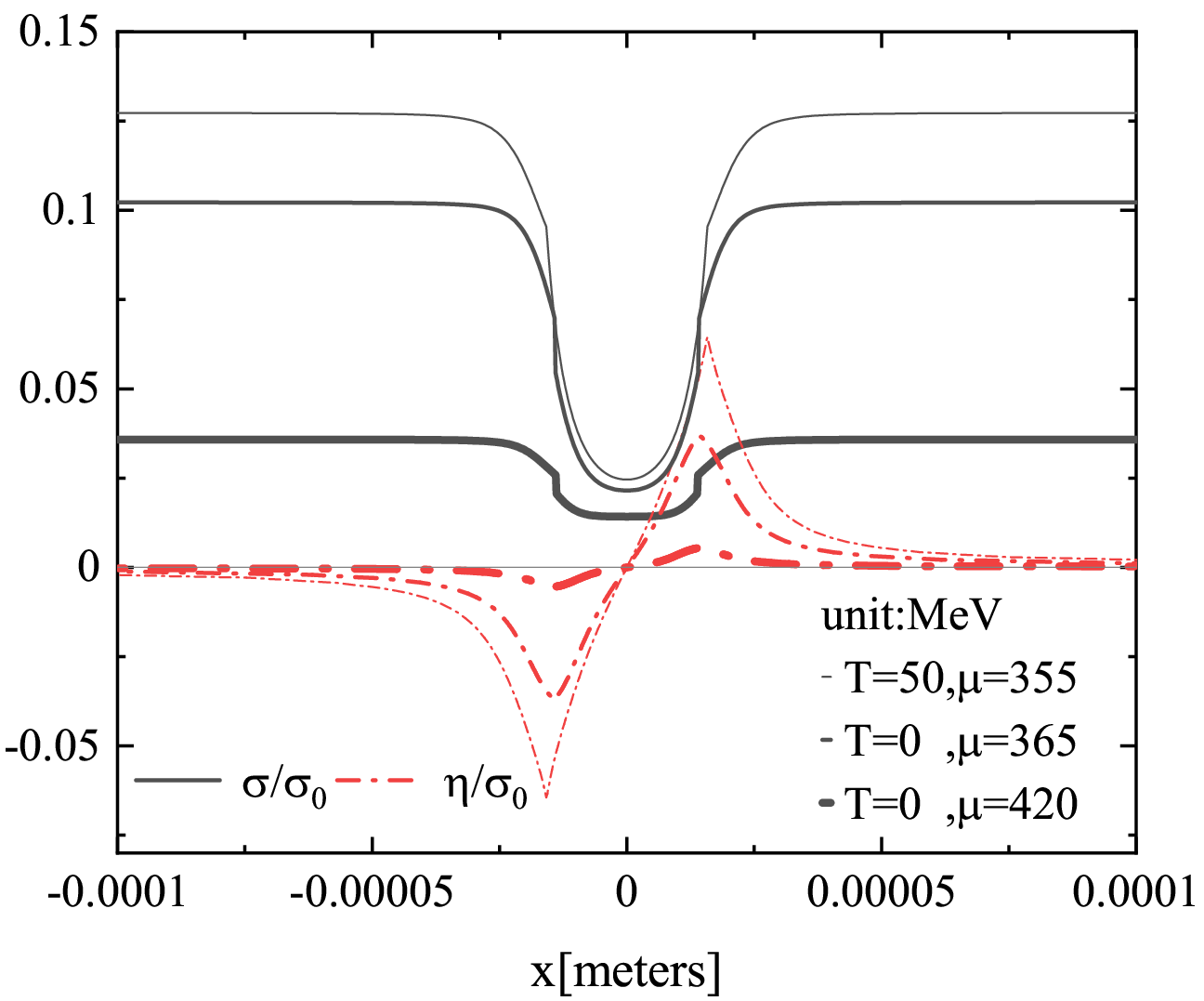}
   \includegraphics[scale=0.35]{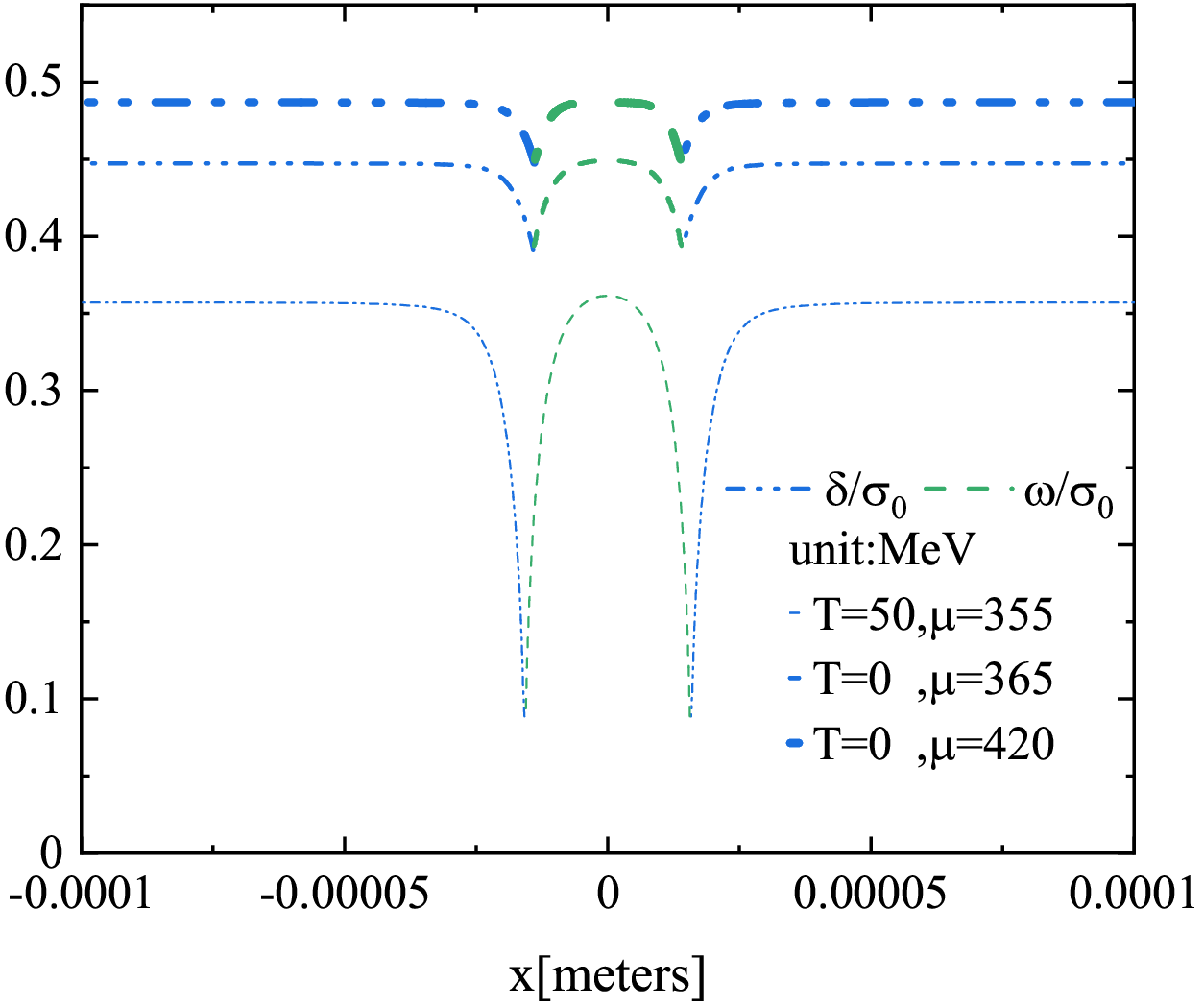}
   \end{center}
   \caption{Axion wall structure: Normalized condensates $\sigma$, $\eta$ (upper), $\delta$, and $\omega$ (lower) 
   in the 2CS quark matter versus $x$ for several sets of $(T,\mu)$ as in Fig.\ref{fg:wallVSx} with $c=0.2$.}
   \label{fg:condVSx}
 \end{figure}

\begin{figure}[ht]
   \centering
   \includegraphics[scale=0.35]{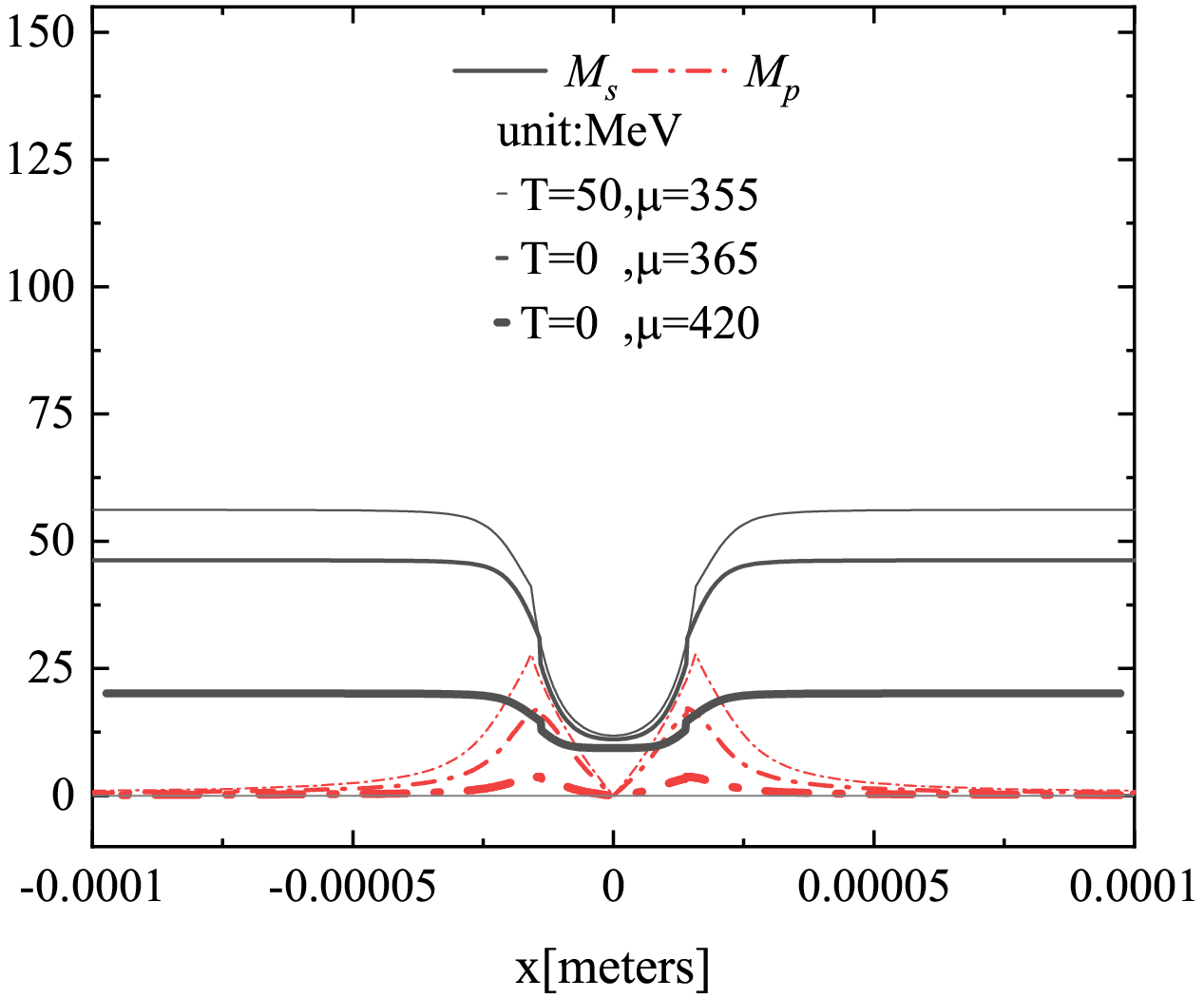}
   \includegraphics[scale=0.35]{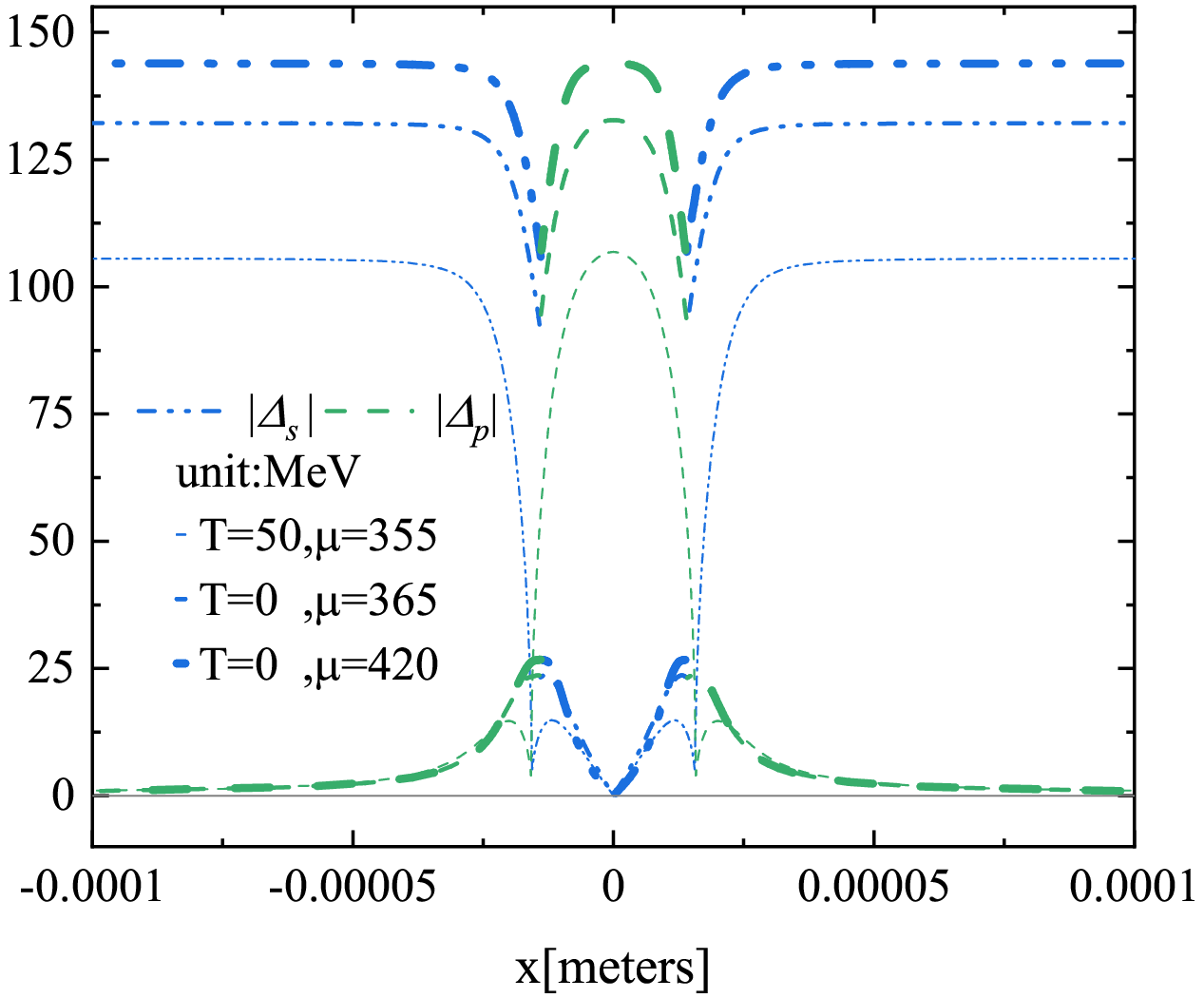}
  \caption{Axion wall structure: Dirac masses $M_s$ and $M_p$ (upper) and Majorana masses $\Delta_s$ and 
  $\Delta_p$ (lower) in the 2CS quark matter versus $x$ for several sets of $(T,\mu)$ as that in Fig.\ref{fg:wallVSx} 
  with $c=0.2$. }
   \label{fg:gapVSx}
 \end{figure} 
 
The structure of the center region of the wall for the CS quark matter, namely the four-type condensates 
and gaps versus $x$ near $x=0$, are shown in Figs.\ref{fg:condVSx} and \ref{fg:gapVSx}, respectively. Same 
to Fig.\ref{fg:wallVSx}, the calculations are performed by fixing $f_a=10^9 \text{GeV}$.  We see that the 
condensates $\omega$ and $\eta$ form near the core of the wall, which indicates the spontaneous breaking of 
the parity symmetry. This region has a distinct boundary at which the condensates $\sigma$, $\eta$, $\delta$, 
and $\omega$ all become discontinuous. In the presence of the 2CS,  $\sigma$ and $\eta$ are suppressed 
significantly and $\omega$ and $\delta$ play dominant roles in the inner and exterior regions of the 
wall, respectively. Analogously, Fig.\ref{fg:gapVSx} shows that in the core (outside core) region, the 
gap $\Delta_p$ ($\Delta_s$) is much larger than the other gaps.             

\begin{figure}[ht]
   \centering
   \includegraphics[scale=0.35]{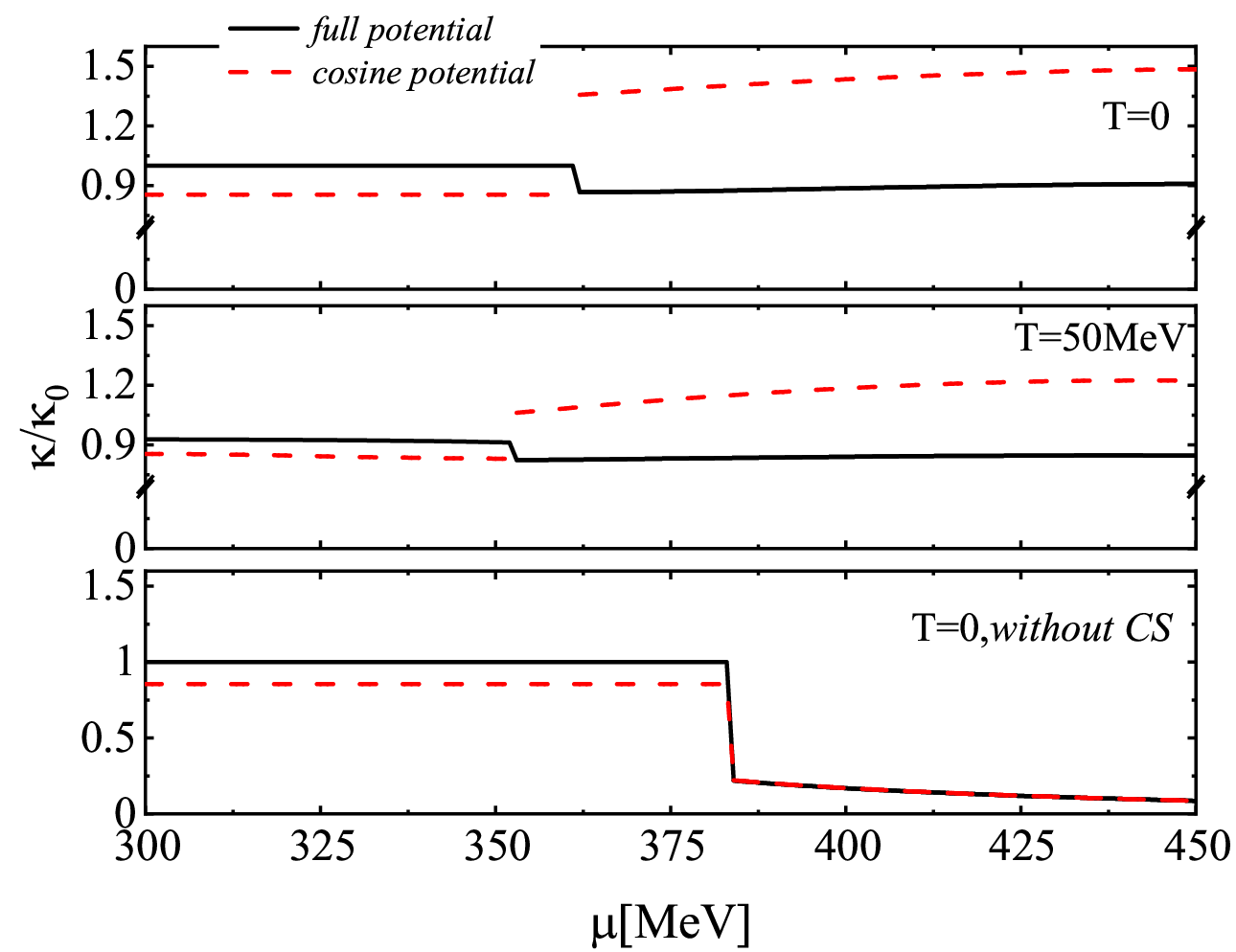}
\caption{Dependence of surface tension $\kappa$ on the quark chemical potential at $T=0$ and $50\,\text{MeV}$. 
For comparison, the result at $T=0$ without considering the CS is also given. This quantity is normalized by 
${\kappa _0} = 1.9 \times {10^{16}}{\rm{Me}}{{\rm{V}}^3}$, the surface tension obtained in vacuum 
for $f_a=10^9 \text{GeV}$. The dash lines are the results calculated using the cosine potential. The parameter 
$c$ is fixed as $0.2$.}
   \label{fg:k}
\end{figure}

The surface tension of the domain wall, namely the energy per unit of transverse area, is defined as
\begin{equation}
	\kappa
	=2\sqrt{2}f_a\int_0^{\pi}d\theta\sqrt{V(\theta)}
	\label{eq:stenion}
\end{equation}
for the case with $v=0$ \cite{GrillidiCortona:2015jxo,Zhang:2023lij}. This quantity at finite $T$ and $\mu$ 
with charge neutrality was first calculated in \cite{Zhang:2023lij}, where the CS was not considered. 
In Fig.\ref{fg:k}, we show $\kappa$ versus $\mu$ at $T=0$ and $50\,\text{MeV}$ in the presence of the CS, which 
is measured by $\kappa_0=1.9 \times 10^{16}\text{MeV}$,  the surface tension obtained at $(T,\mu)$=$(0,0)$ for 
$f_a=10^9 \text{GeV}$. We see that for $T=0$,  $\kappa/\kappa_0$ decreases from one to $\sim 0.85$ at the critical 
chemical potential and then grows slowly with $\mu$. This contrasts with the case without considering the CS 
where $\kappa/\kappa_0$ drops abruptly from one to $\sim 0.20$ at the phase transition point, as shown 
by the solid line. For $T=50\,\text{MeV}$, the surface tension is weakened in both the chiral symmetry breaking 
and CS phases but is still considerable compared to its value at zero $T$. In addition, Fig.\ref{fg:k} shows that 
the analytic formula \begin{equation}
	\kappa = 
	8m_a f_a^2 =\frac{8\chi_t}{m_a},
	\label{eq:k-analytic}
\end{equation}
which is obtained by using the simple cosine potential $V(\theta)=m_a^2f_a^2(1-\cos\theta)$ in \cite{Zhang:2023lij},  
does not work in the presence of the CS, even it is a very good approximation at larger $T/\mu$ for the case without 
the CS \cite{Zhang:2023lij}.      

It has been argued in \cite{Zhang:2023lij} that forming axion domain walls in the bulk quark matter cost 
zero energy in the thermodynamic limit. Namely, forming axion walls in bulk quark matter is more easier than 
forming axion walls in the vacuum. So it might be possible that the axion walls are abundant in the cores 
of neutron stars. The argument also holds if the bulk quark matter is in the 2CS phase.    

\section{ \label{sec:conclusion} Conclusion and Outlook }

We study the QCD axion potential in dense quark matter by simultaneously taking into account the scalar 
and pseudo-scalar condensates in both the quark-antiquark and diquark channels. We employ the two flavor 
NJL model with two types of four-quark interactions: one arising from single-gluon exchange and another 
induced by the instantons. By performing the Fierz transformation, the QCD axion field can be introduced 
through the instanton induced interactions in both the quark-antiquark and diquark channels in this model. 
We first obtain the analytic dispersion relations of quarks in the presence of axion field at the mean field 
level, which involves two scalar condensates $\sigma$ and $\delta$ and two pseudo-scalar condensates $\eta$ 
and $\omega$. We then calculate the axion potential, the axion mass (or topological susceptibility), the 
quartic self-coupling, and the domain wall tension at finite $T/\mu$. We mainly focus on the influences 
of the chiral phase transition on these quantities and the effects of Dirac-type masses in the presence 
of the 2CS. 

We found that for larger $\mu$ and lower $T$, the two diquark condensates $\delta$ and $\omega$ 
can't exist simultaneously: the former emerges in the ranges $\theta=[0,\pi/2)$ and $(3\pi/2,2\pi]$, 
while the later in the range $\theta=(\pi/2,3\pi/2)$. Namely, there is a phase transitions at 
$\theta=\pi/2\,(3\pi/2)$ where $\delta\,(\omega)$ drops suddenly to zero. This is consistent with the 
massless case considered in \cite{Murgana:2024djt}. In contrast, $\sigma$ and $\eta$ can form simultaneously in the 
CS phase except at $\theta=0$ and $\pi$ where $\eta$ vanishes. We confirmed that the periodicity of the 
axion potential $V(\theta)$ with a period $\pi$ found in \cite{Murgana:2024djt} is disrupted due to 
the nonzero Dirac masses. However, the axion potential still exhibits double peaks at $\theta=\pi/2$ 
and $3\pi/2$ and a local minimum at $\theta=\pi$ in the presence of the 2CS: for large enough $\mu$,  $V(\pi)$ 
is approaching $V(0)$ and thus $\pi$ can still be regarded as a good approximate period. 

We concluded that the chiral restoration transition does't always lead to a reduction of $m_a$. 
Instead, the chiral transition with the emergence of the 2CS results in an abrupt increase of 
$m_a$ for the parameter range $c\approx(0.05,0.45)$ at $T=0$. The same conclusion holds for the 
topological susceptibility since it is proportional to $m_a^2$. In addition, the sign and strength 
of the quartic self-coupling $\lambda_a$ in the CS phase are quite sensitive to $c$.
In most of the range $c=(0,0.5)$, the strength of $\lambda_a$ is enhanced significantly by the chiral 
phase transition due to the appearance of 2CS. Moreover, the $\mu$-dependence of $|\lambda_a|$ is 
also quite sensitive to $c$. 
    
We confirmed that the width of the domain wall still keeps quite narrow in the chiral symmetric 
phase with the 2CS at moderate and high $\mu$. Unlike the axion mass, the wall tension in the 2CS 
phase reduces slightly compared to it's value in the chiral symmetry breaking phase at $T=0$ for 
$c=0.2$. This is very different from the case without the 2CS where the wall tension decreases significantly 
at the chiral phase transition point. We found that the wall tension obtained from the full potential with 
the 2CS deviates greatly from that calculated using the simple cosine potential.                            

In this paper, we don't consider the charge neutrality and $\beta$-equilibrium constraints which 
must be taken into account for compact stellar objects such as neutron and protoneutron stars. 
Our work along this direction is in progress. In addition, it is interesting to investigate the more 
physical situation with 2+1 flavors by including the strange quark at moderate and high chemical 
potentials. In this case, more condensates need to be included and the six-fermion interaction 
induced by the instantons should play a very important role. Furthermore, it is deserved to study 
the axion effect on the properties of the neutron and protoneutron star where the CS phase may 
appear in their cores. We leave these topics to future works.              

\vspace{5pt}
\noindent{\textbf{\large{Acknowledgements}}}\\\\
This work was supported by the National Natural Science Foundation of China (NSFC) under Grant No. 11875127.

\appendix
\vspace{2pt}

\section{ Topological susceptibility and axion self-coupling for different ratios of  $H_1/G_1$ and $H_2/G_2$}

Here we report the numerical calculations on the topological susceptibility (axion mass) and axion self-coupling 
using the ratios of $H_1/G_1$ and $H_2/G_2$ beyond the Fierz transformations.    

\begin{figure}[ht]
   \centering
   \includegraphics[scale=0.35]{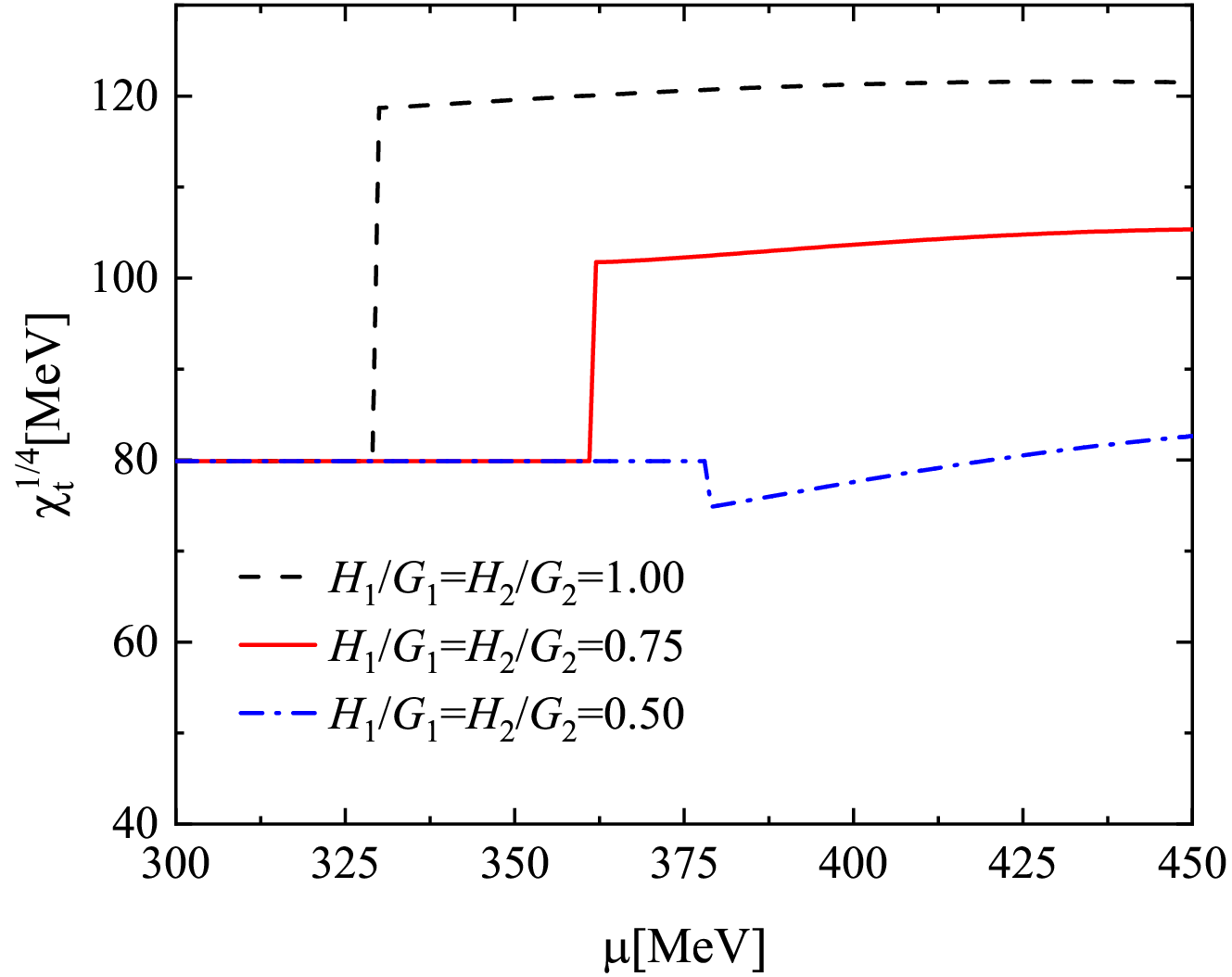}
\caption{ Topological susceptibility $\chi^{1/4}_t$ versus $\mu$ for different ratios of $H_1/G_1$ 
and $H_2/G_2$ at $T=0$. For simplicity, $H_1/G_1$=$H_2/G_2$ is assumed. The parameter $c$ is fixed as $0.2$. }
   \label{fg:H12top}
\end{figure}


\begin{figure}[ht]
   \centering
   \includegraphics[scale=0.35]{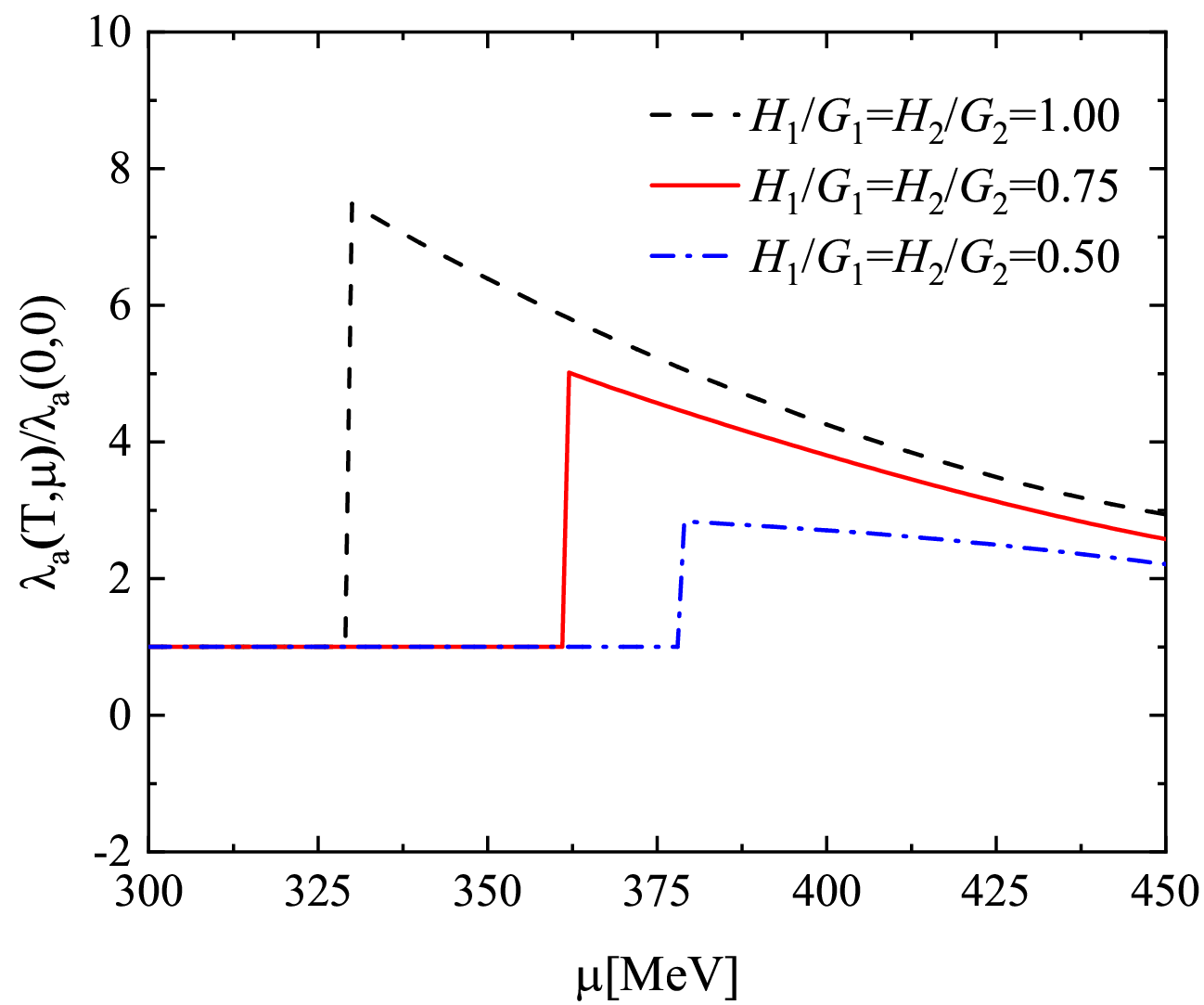}
\caption{Normalized axion self-coupling $\lambda_a$ versus $\mu$ under the same conditions as that in Fig.\ref{fg:H12top}}
   \label{fg:H12selfcoupling}
\end{figure}

Figure \ref{fg:H12top} shows the topological susceptibility $\chi^{1/4}_t$ versus $\mu$ for different values of 
$H_1/G_1$ and $H_2/G_2$ at $T=0$. The parameter $c$ is fixed as $0.2$ and $H_1/G_1=H_2/G_2=r$ is assumed for simplicity. 
Fig.\ref{fg:H12top} indicates that $\chi^{1/4}_t$ decreases with the decrease of $r$ in the 2CS phase. For weaker 
diquark interactions with $r=0.5$, $\chi_t$ drops at the chiral transition point. But such a decrease is not so significant 
due to the presence of the 2CS (the gap $\Delta_s$ is $\sim 70\,\text{MeV}$ at $(T,\mu)=(0,400)$ in this case).      

Figure \ref{fg:H12selfcoupling} displays the axion self-coupling $\lambda_a$ versus $\mu$ under the same conditions 
as that in Fig.\ref{fg:H12top}. Due to the appearance of the 2CS, the self-coupling gets enhanced at the chiral transition 
point for all the cases: the stronger the diquark couplings, the more sharply the self-coupling increases at the critical point.
Especially for $r=0.5$, even $\chi_t$ declines at the phase transition point, $\lambda_a$ still increases near threefold.

\begin{figure}[ht]
   \centering
   \includegraphics[scale=0.35]{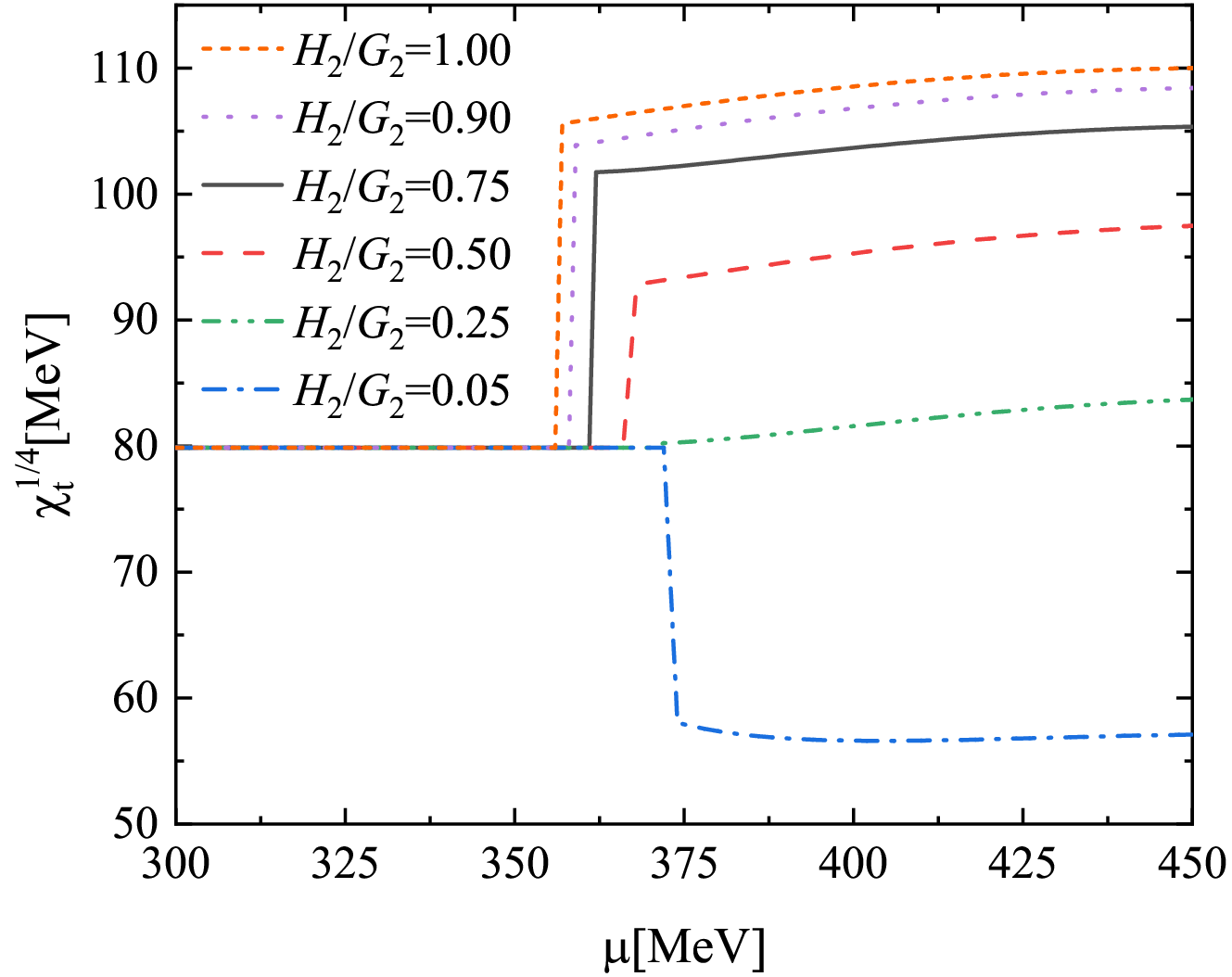}
\caption{ Topological susceptibility $\chi^{1/4}_t$ versus $\mu$ for different ratios of $H_2/G_2$ at $T=0$. 
 The parameter $c$ is fixed as $0.2$ and $H_1/G_1$=0.75 is fixed by the Fierz transformation.}
   \label{fg:H2top}
\end{figure}

\begin{figure}[ht]
   \centering
   \includegraphics[scale=0.35]{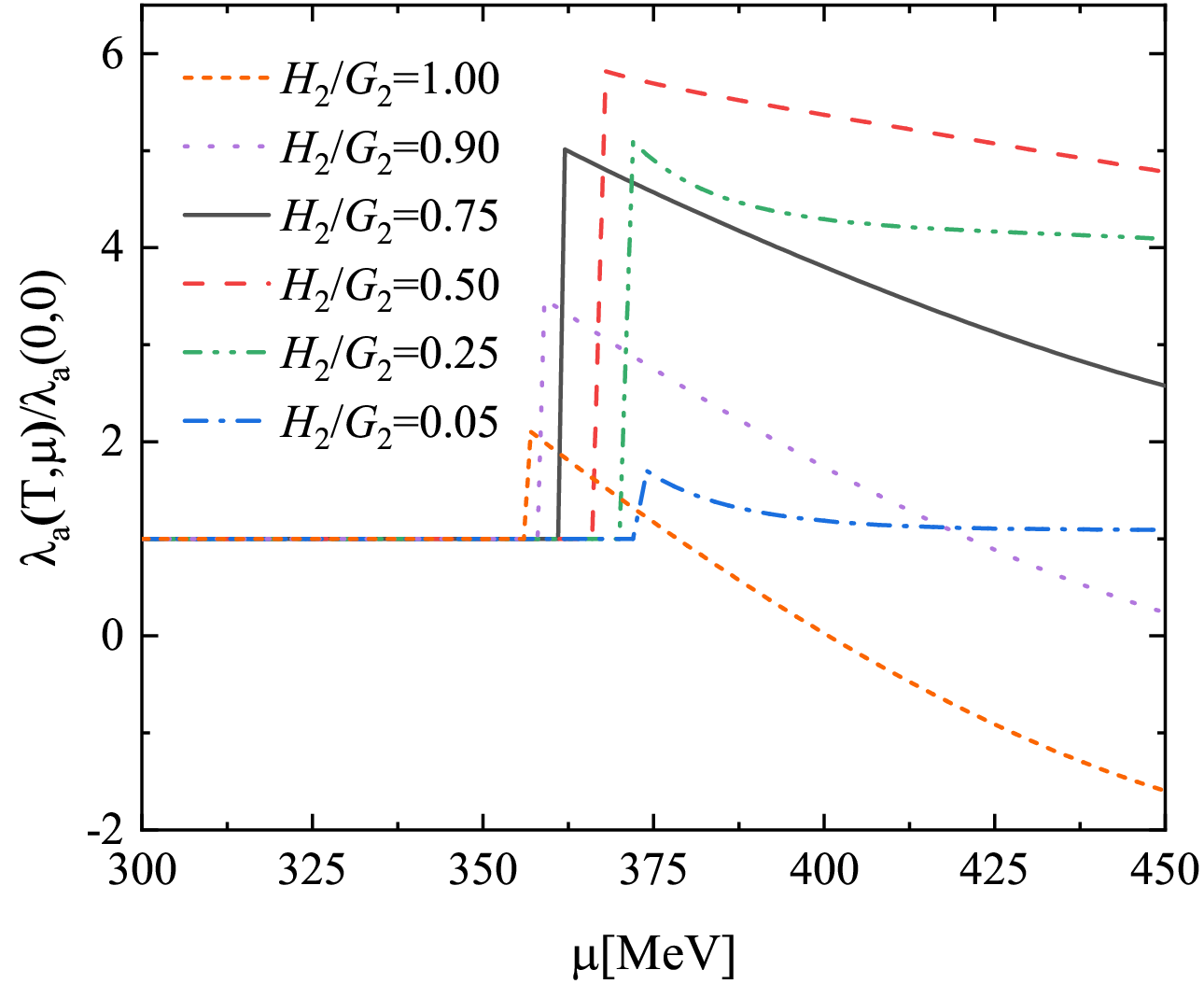}
\caption{Normalized axion self-coupling $\lambda_a$ versus $\mu$ under the same conditions as that in Fig.\ref{fg:H2top}}
   \label{fg:H2selfcoupling}
\end{figure}

We also show $\chi_t$ and $\lambda_a$ versus $\mu$ for different ratios of $H_2/G_2$ at $T=0$ in Figs.\ref{fg:H2top}
and \ref{fg:H2selfcoupling}, respectively, where the parameter $c$ is fixed as $0.2$ and $H_1/G_1$=0.75 is fixed by 
the Fierz transformation.    

\newpage


\end{document}